\documentclass[12pt,oneside]{revtex4}
\usepackage{amssymb}
\usepackage{footmisc}
\usepackage{natbib}
\usepackage{anysize}
\usepackage{graphicx}
\usepackage{amsmath}
\usepackage{wasysym}
\newcommand{\xiaowu}{\fontsize{10.0pt}{\baselineskip}\selectfont}
\usepackage{setspace}
\usepackage[bookmarks = false]{hyperref}
\usepackage{color}
\usepackage{subfigure}
\usepackage{multirow}
\usepackage[T1]{fontenc}
\renewcommand{\baselinestretch}{1.5}
\renewcommand{\andname}{\ignorespaces}
\marginsize{2.3cm}{2.3cm}{1.5cm}{1.5cm}
\usepackage{color}
\definecolor{SH}{RGB}{0,0,200}
\definecolor{SH2}{RGB}{0,200,200}
\begin{document}

\title{Quantum storage of 1650 modes of single photons at telecom wavelength}

\author{Shi-Hai Wei$^{1}$}
\author{Bo Jing$^{1,\dagger}$}
\author{Xue-Ying Zhang$^{1}$}
\author{Jin-Yu Liao$^{1}$}
\author{Hao Li$^{4}$}
\author{Li-Xing You$^{4}$}
\author{Zhen Wang$^{4}$}
\author{You Wang$^{1,3}$}
\author{Guang-Wei Deng$^{1,2}$}
\author{Hai-Zhi Song$^{1,3}$}
\author{Daniel Oblak$^{5,\ddagger}$}
\author{Guang-Can Guo$^{1,2}$}
\author{Qiang Zhou$^{1,2,\ast}$}

\affiliation{$^1$Institute of Fundamental and Frontier Sciences $\&$ School of Optoelectronic Science and Engineering, University of Electronic Science and Technology of China, Chengdu 610054, P. R. China}
\affiliation{$^2$CAS Key Laboratory of Quantum Information, University of Science and Technology of China, Hefei 230026, P. R. China}
\affiliation{$^3$Southwest Institute of Technical Physics, Chengdu 610041, P. R. China}
\affiliation{$^4$Shanghai Institute of Microsystem and Information Technology, Chinese Academy of Sciences, Shanghai 200050, P. R. China}
\affiliation{$^5$Institute for Quantum Science and Technology, and Department of Physics $\&$ Astronomy, University of Calgary, 2500 University Dr. NW, Calgary, Alberta T2N 1N4, Canada}
\affiliation{Correspondence and requests for materials should be addressed to BJ (email: $^{\dagger}$bjing@uestc.edu.cn), DO (email:  $^{\ddagger}$doblak@ucalgary.ca), or QZ (email: $^{\ast}$zhouqiang@uestc.edu.cn).}

\maketitle


\textbf{
	\\To advance the full potential of quantum networks one should be able to distribute quantum resources over long distances at appreciable rates. As a consequence, all components in the networks need to have large multimode capacity to manipulate photonic quantum states. Towards this end, a multimode photonic quantum memory, especially one operating at telecom wavelength, remains a key challenge. Here we demonstrate a spectro-temporally multiplexed quantum memory at 1532 nm. Multimode quantum storage of telecom-band heralded single photons is realized by employing the atomic frequency comb protocol in a 10-m-long cryogenically cooled erbium doped silica fibre. The multiplexing encompasses five spectral channels - each 10 GHz wide - and in each of these up to 330 temporal modes, resulting in the simultaneous storage of 1650 modes of single photons. Our demonstrations open doors for high-rate quantum networks, which are essential for future quantum internet.}

\vspace{0.5cm}


\renewcommand\section[1]{
	\textbf{#1}
}
\section{\\Introduction}
\\Multimode capacity is an essential requirement to achieve high data rates in modern communication networks. Towards a future quantum network\cite{kimble2008quantum,simon2017towards,wehner2018quantum,Wei2022} compatible with existing telecom infrastructure, this requirement must also be applied\cite{jin2015telecom,saglamyurek2015quantum}. One challenge in pursuing such a quantum network is to develop a multimode quantum memory\cite{collins2007multiplexed,simon2007quantum,lvovsky2009optical,simon2010quantum,heshami2016quantum}, which is able to simultaneously store and process multiple modes of single photons in various degrees of freedom, such as in temporal degree\cite{de2008solid,usmani2010mapping,bonarota2011highly,hosseini2011high,gundougan2013coherent,tang2015storage,jobez2016towards,tiranov2016temporal,laplane2017multimode,wen2019multiplexed,heller2020cold,liu2021heralded,lago2021telecom,su2022demand,ortu2022multimode}, spectral degree\cite{sinclair2014spectral,saglamyurek2016multiplexed,askarani2021long}, spatial degree\cite{lan2009multiplexed,grodecka2012high,nicolas2014quantum,ding2015quantum,zhou2015quantum,parniak2017wavevector,pu2017experimental,chrapkiewicz2017high,tian2017spatial}, or any combination of these\cite{yang2018multiplexed,seri2019quantum}. In addition to the usual figures-of-merit for quantum memories, i.e., efficiency, time, and fidelity, the multimode performance of a quantum memory is also important, which is mainly determined by the number of storage channels, storage time, and storage bandwidth. The storage channel could be implemented in the spectral or spatial domain. The storage time and storage bandwidth of each channel determine an upper bound of temporal mode number in each channel.

On the quest to reach large multimode quantum storage capacity, the use of multiple spatial channels has yielded good results, with a channel number up to 665 realized in gaseous atomic ensembles\cite{chrapkiewicz2017high} based on Duan–Lukin–Cirac–Zoller (DLCZ) protocol\cite{duan2001long}. Further improvements to the multimode capacity of such memory could be achieved via increasing the number of stored temporal modes in each of these channels. However, the simultaneous storage of multiple temporal modes remains challenging in atomic gas ensembles due to limitations of the applied storage protocols \cite{duan2001long,turukhin2001observation,reim2010towards}. Specifically, the number of temporal modes that can be stored is related to the optical depth (OD) of the storage media\cite{nunn2008multimode}. Fortunately, the atomic frequency comb (AFC) quantum memory with rare-earth ion-doped (REID) materials does not feature any constraint in terms of OD for simultaneous temporal storage, and, thus, is promising for developing highly multimode photonic quantum memory using several photonic degrees of freedom\cite{afzelius2009multimode}. Multiple storage channels have been demonstrated with the AFC protocol in the spatial domain\cite{zhou2015quantum} and in the spectral domain\cite{sinclair2014spectral,saglamyurek2016multiplexed,askarani2021long}, which is facilitated by the inhomogeneous broadening in REID materials. Storage of multiple temporal modes in one spectral channel has also been achieved, for instance the storage of 1250 temporal modes in Yb$^{3+}$:Y$_2$SiO$_5$ at 979 nm\cite{businger2022non}. Furthermore, the storage of multiple temporal modes in multiple channels has been reported, including storage of 12 spectro-spatial-temporal modes\cite{yang2018multiplexed} and 130 spectro-temporal modes\cite{seri2019quantum} in Pr$^{3+}$:Y$_2$SiO$_5$ at 606 nm. Despite these important advances, a quantum memory with large multimode capacity at telecom wavelength has yet to be demonstrated, as an essential step towards a future quantum network compatible with existing telecom infrastructure.

In this paper, we present the multimode storage of 1650 modes of single photons at telecom wavelength with a 10-m-long cryogenically cooled erbium doped silica fibre (EDF). In our demonstration, five individual AFC spectral channels - each with 10 GHz bandwidth and separated by 5 GHz isolation - are prepared by using an optical frequency comb\cite{fortier201920} combined with frequency chirping \cite{saglamyurek2016multiplexed}. Up to 330 temporal modes of heralded single photons generated from cascaded second-order nonlinear process\cite{lefebvre2021compact,zhang2021high,yu2022spectrally} are stored in each spectral channel. Our achievements pave the way towards developing future quantum internet by utilizing multimode quantum memory compatible with the infrastructures of fibre-based communication.

\section{\\Results}
\begin{figure}[ht]
\centering
\includegraphics[]{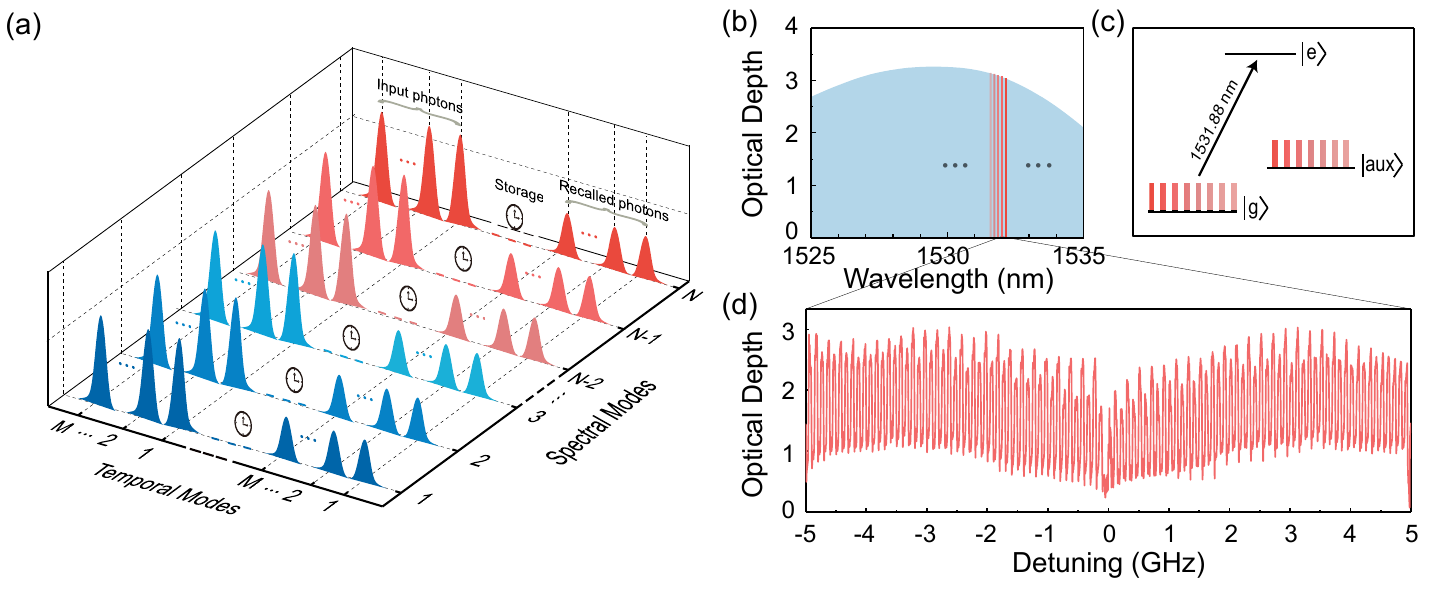}
\caption{\xiaowu\textbf{Spectro-temporal multimode quantum storage of single photons at telecom wavelength.} (a) General scheme for storage of N $\times$ M modes of single photons. A train of M temporal modes are stored into the AFCs with N spectral channels, resulting in storage of N $\times$ M modes of single photons. (b) The absorption profile of $^{4}$I$_{15/2}$ to $^{4}$I$_{13/2}$ transitions of Er$^{3+}$ ions in the EDF at 10 mK. The inhomogeneous broadening is up to 2 THz (only 1~THz are shown here). In the section of inhomogeneous broadening absorption profile (70~GHz wide), we prepare five 10-GHz-wide AFCs with a separation of 5~GHz between the edges of adjacent AFCs. In principle, over seventy such AFCs can be prepared if we make use of the entire inhomogeneous broadening. (c) Simplified energy level scheme of Er$^{3+}$ ions in erbium doped silica fibre. AFCs are prepared through frequency-selective optical pumping that transfers atomic states from the ground state ($\vert$g$\rangle$) to the auxiliary state ($\vert$aux$\rangle$) via the excited state ($\vert$e$\rangle$), forming AFCs. (d) A typical trace of 10-GHz-wide AFC measured in our experiment (the central wavelength of the AFC is 1532.00 nm, and comb spacing $\Delta$ is 100 MHz, leading to T$_{storage}$=10 ns).}
\label{fig:1}
\end{figure}

\begin{figure}[ht]
\centering
\includegraphics[width=16cm]{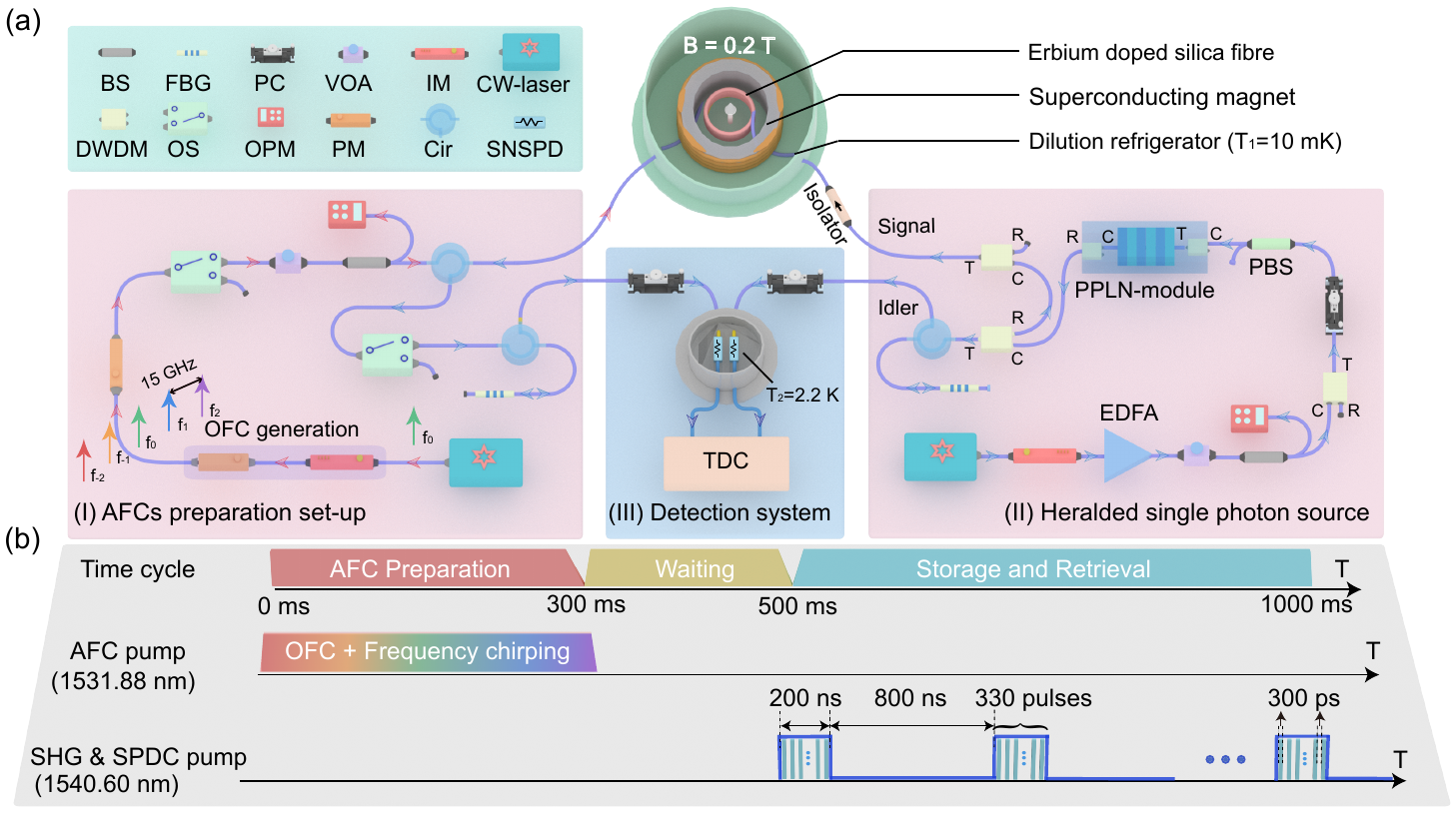}

\caption{\xiaowu\textbf{Experimental setup and time sequences.} (a) Experimental setup. (I) Preparation of AFC memory with five spectral channels. The erbium doped silica fibre (EDF) is cooled to a temperature of 10 mK by a dilution refrigerator and exposed to a magnetic field of 2000 Gauss provided by a superconducting magnet. Light from a continuous-wave laser (CW-laser) is firstly modulated to an optical frequency comb (OFC) by an intensity modulator (IM) and a phase modulator (PM). Then another PM is utilized to generate chirped light on each comb of the OFC. The modulated pump light is sent to the EDF via an optical circulator (Cir). Optical switch (OS) and variable optical attenuator (VOA) are used to control the pump time and intensity. Beam splitter (BS) and optical power meter (OPM) are utilized to monitor the pump light power. Fibre Bragg grating (FBG) is used to select signal photons recalled from different AFCs. (II) Preparation of the heralded single photon source (HSPS). Light from a CW-laser is modulated to 300~ps pulses by an IM and sent to the periodically poled lithium niobate (PPLN) module for generating correlated photon-pairs. By utilizing two dense wavelength division multiplexers (DWDMs), signal photons at 1531.88~nm and idler photons at 1549.32~nm with a bandwidth of 100~GHz are filtered out. FBG is used to select the idler photons corresponding to different spectral channels. Subsequently idler photons are directly sent to the detection system. The signal photons are sent into the EDF for storage. Erbium doped fibre amplifier (EDFA) and VOA are used to adjust the pump power. Polarization controller (PC) and polarizing beam splitter (PBS) are utilized to control the polarization of pump light. (III) Detection system. The idler and recalled signal photons are detected by two superconducting nanowire single photon detectors (SNSPDs). Coincidence measurements are performed by a time-to-digital converter (TDC). PCs are used to control the polarizations of signal and idler photons. (b) Time sequences. The cycle time of the experiment is one second, including 300 ms for AFCs preparation, 200 ms for waiting spontaneously emitted photons from excited states, and 500 ms for storage. The AFC pump laser is modulated to OFC and frequency chirped light within the preparation time of 300 ms. The SHG \& SPDC pump laser with the repetition of 1 $\mu$s is modulated into 330 light pulses per cycle with pulse durations of and spaced by 300 ps (more details see Methods).}
\label{fig:2}
\end{figure}

\\
\textbf{Multimode quantum storage scheme.} Based on the inhomogeneous broadening in REID materials, an N $\times$ M multimode quantum memory can be realized by using N spectral channels and M temporal modes in each spectral channel (see Fig. \ref{fig:1}(a)). Through spectral tailoring of the inhomogeneous broadened absorption line of REID materials, AFC channels with a number of N in the spectral domain are prepared. In each spectral channel, a train of single photons in M temporal modes are stored, with the maximum value of M determined by the time-bandwidth product of the AFC memory. By doing so, a total number of N $\times$ M modes of single photons are stored, simultaneously. To develop such a quantum memory with multimode capacity at telecom wavelength, a promising REID material is low doping concentration EDF, which has a telecom-band transition wavelength and THz-wide inhomogeneous broadening (see Figs. \ref{fig:1}(b), (c)).
\\
\textbf{Experimental setup.} The experimental setup is composed of an EDF based AFC quantum memory with five spectral channels, a heralded single photon source at telecom wavelength, and a coincidence detection system, as illustrated in Fig. \ref{fig:2}(a). The experimental time sequences are also shown in Fig. \ref{fig:2}(b) (see Methods).

We prepare an AFC quantum memory with five spectral channels in a 10-m-long EDF cooled to a temperature of 10 mK and exposed to a magnetic field of 2000 Gauss (see Fig. \ref{fig:2}(a)(I) and Methods for details). More precisely, we propose and apply an approach - optical frequency comb combined with frequency chirping - to prepare five individual AFCs, each with a bandwidth of 10 GHz, at central wavelengths of 1532.11 nm, 1532.00 nm, 1531.88 nm, 1531.76 nm and 1531.65 nm, labeled as channels of 1, 2, 3, 4, and 5 (see Supplementary Information Note 1 for more details on preparing five AFCs by using optical frequency comb combined with frequency chirping method). Figure \ref{fig:1}(d) depicts a typical AFC (channel 2) in our experiment, and all the five AFCs are shown in Supplementary Information Note 1. Note that different from using multiple individual lasers in parallel to prepare spectrally multiplexed AFC quantum memory\cite{saglamyurek2016multiplexed}, our method not only prepares such an AFC quantum memory by using a single laser, but also guarantees each channel with large bandwidth. Moreover, the scheme applies spectral isolation among channels, thus making the crosstalk among different storage channels to be negligible. When five spectral distinct photons - spectra matching with the five AFCs - are absorbed by the five AFCs respectively, five collective atomic excitation states can be generated. The state in one of the channels, e.g., in channel 1, can be expressed as:
\begin{equation}
|\Psi_{1}\rangle=\frac{1}{\sqrt{W_1}}\sum_{j=1}^{W_{1}}c_{j}e^{i2\pi\delta_{j}t}e^{-ikz_{j}}|g_{1}g_{2}...e_{j}...g_{W_1}\rangle,
\end{equation}
where $W_1$ is the total atom number constituting the AFC in channel 1, the amplitude $c_{j}$ depends on both the resonance frequency and position of the $j_{th}$ atom, $\delta_{j}$ is the detuning of the $j_{th}$ atom with respect to the photon carrier frequency, $k$ is the wave number of the input photon, $z_{j}$ is the position of the $j_{th}$ atom, $|g_{j}\rangle$ and $|e_{j}\rangle$ represent ground and excited states of the $j_{th}$ atom, respectively. Following the creation of the collective atomic excitation, the different terms in the collective excitation state $|\Psi_1\rangle$, having different detunings, begin to accumulate different phases. Due to the periodic nature of AFC, i.e., $\delta_{j}=m_{j}\Delta$ ($m_j=0,\pm 1, \pm 2,...$ and $\Delta$ is the teeth spacing of AFC), at the time $t=1/\Delta$ all terms in the state acquire phases equal to an integer multiple of 2$\pi$, which are all equivalent to 0. This process of rephasing leads to the re-emission of input photon in its original quantum state.

To characterize the five spectral channels of quantum memory, we generate heralded single photons at telecom wavelength in a fibre pig-tailed periodically poled lithium niobate (PPLN) module\cite{lefebvre2021compact,zhang2021high,yu2022spectrally}, as shown in Fig. \ref{fig:2}(a)(II). The light from a continuous-wave (CW) laser (PPCL300, PURE Photonics) operating at 1540.60 nm is modulated to light pulses and subsequently sent into the PPLN module. The cascaded second-harmonic generation (SHG) and spontaneous parametric down conversion (SPDC) processes result in correlated photon-pairs centred at 1540.60 nm with a bandwidth of $\sim$60 nm. After efficient filtering, the central wavelengths of signal and idler photons are 1531.88 nm and 1549.32 nm with a bandwidth of $\sim$100 GHz. The signal photons are sent into the five AFC spectral channels. It is worth noting that AFC itself is both a memory and a filter\cite{saglamyurek2014integrated}. In our experiment, the spectral modes that do not match the storage bandwidth will be either absorbed and re-emitted spontaneously (with a decay time of $\sim$10 ms) or pass through the memory, in which case they can be discriminated by their arrival time. Then, recalled signal photons from different AFCs are selected by a tunable fibre Bragg grating (FBG). The idler photons corresponding to different recalled signal photons are further selected by another FBG. The two FBGs enable the selective detection of different frequency modes employed in spectral multiplexing (see Supplementary Information Note 2 for more details of heralded single photon source).

Finally, we establish a coincidence detection system consisting of superconducting nanowire single photon detectors (SNSPDs, P-CS-6, PHOTEC Corp.) and a time-to-digital converter (TDC, quTAG, qutools), as shown in Fig. \ref{fig:2}(a)(III) (see Methods for details of the SNSPDs). All SNSPD detection signals are delivered to the TDC to perform coincidence measurements. The system storage efficiency can be calculated by the ratio of the counts of recalled heralded photons to that of the input heralded photons. The second-order cross-correlation function ($g_{s,i}^{(2)}(0)$) between signal and idler photons after/before storage is calculated as:
\begin{equation}
g_{s,i}^{(2)}(0)=\frac{p_{si}}{p_s\cdot p_i},
\end{equation}
where $p_{si}$ is the probability of 3-fold coincidence detections of trigger signal, idler photons, and signal photons, $p_s$ ($p_i$) is the probability of coincidence detections of trigger signal and signal (idler) photons, respectively. Note that the coincidence detection in our experiment with trigger signal enables us to select the recalled photons by their arrival time, thus reducing the noise photon from the residual spontaneous emission. According to the Cauchy-Schwarz inequality\cite{chou2004single}, a non-classical field satisfies $g_{s,i}^{(2)}(0)>2$. More details of measuring $g_{s,i}^{(2)}(0)$ through coincidence counts are given in Methods.
\\
\textbf{Measurement.} First, we investigate the performance of the five AFC spectral channels. To this end, we vary the storage time from 5 ns to 230 ns by programming the tooth spacing from 200 MHz to 4.35 MHz. For each case, we map heralded single photons onto each of the five spectral channels, and subsequently collect coincidence statistics of the recalled photons to gauge storage performance in terms of storage time, efficiency, and preservation of quantum properties manifested in the $g_{s,i}^{(2)}(0)$ values. According to the measured results, the system storage efficiency of each channel with different storage times ranges from 0.1\% to 1\%. The $g_{s,i}^{(2)}(0)$ still remains above 25 for the maximum storage time of 230 ns (see Table I), demonstrating strong non-classical properties for all storage times (see Supplementary Information Note 3 and 4 for more analysis of the storage efficiency and $g_{s,i}^{(2)}(0)$ for five individual AFCs). Furthermore, these measurements demonstrate that our memory, in principle, is able to perform quantum storage of 2300 (10 GHz $\times$ 230 ns) temporal modes of telecom-band single photons in each channel, for five parallel channels amounting to a total of 11500 modes.

\begin{figure}[ht]
\centering
\includegraphics[]{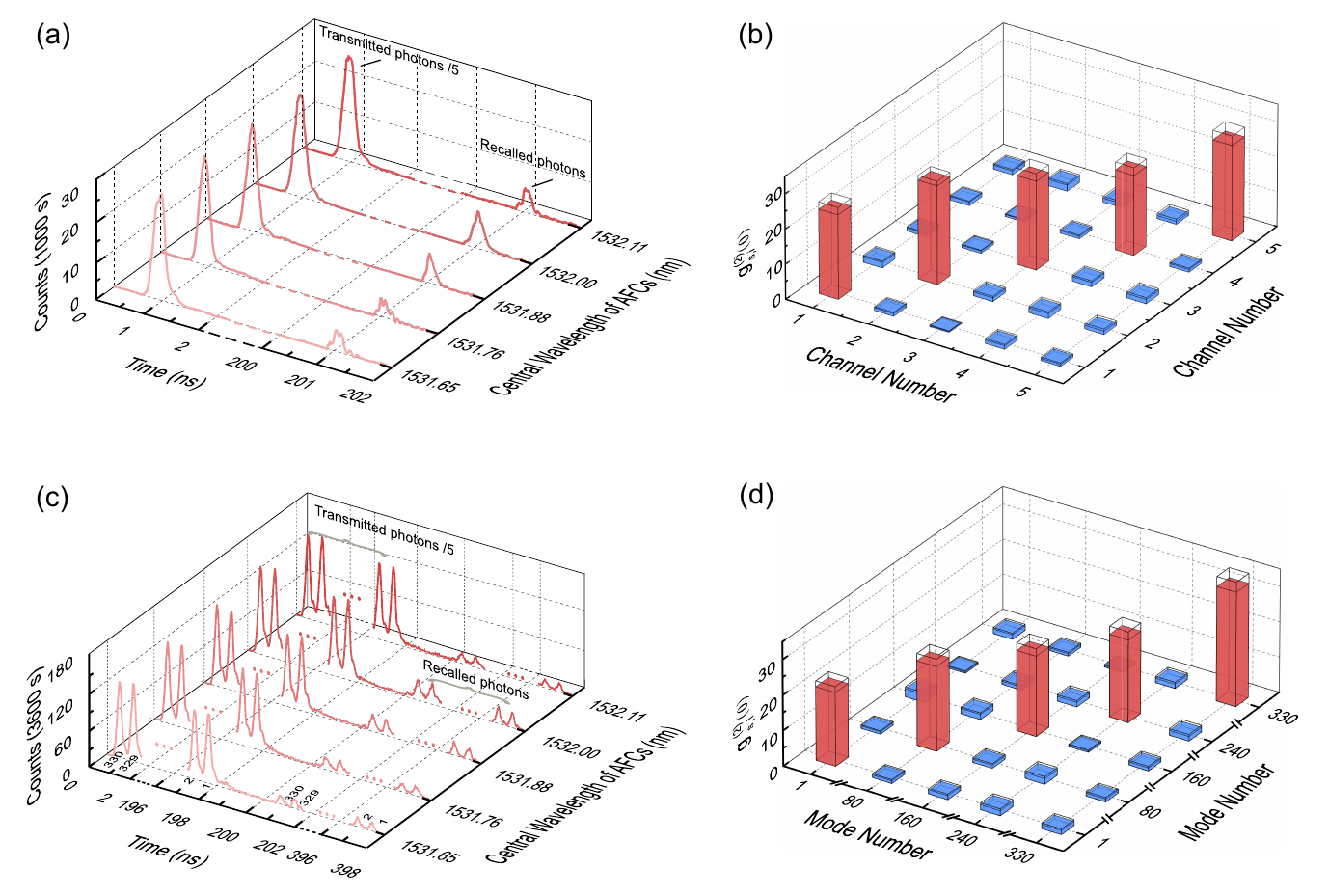}
\caption{\xiaowu\textbf{Characterization for the multimode quantum memory.} (a) Results for spectral-multiplexed storage in five AFC spectral channels with a storage time of 200 ns. (b) Crosstalk between different spectral modes of idler photons and recalled photons. The average measured $g_{s,i}^{(2)}(0)$ between recalled photons and idler photons is 25.48 $\pm$ 1.05 for 5 correlated spectral modes and 1.01 $\pm$ 0.10 for 20 uncorrelated modes. (c) Results for storage of 1650 spectro-temporal modes of single photons at telecom wavelength. Heralded single photons with 330 temporal modes are generated and stored into five spectral channels with a storage time of 200 ns, totally, 1650 modes are stored into the five AFCs. (d) Crosstalk between different temporal modes of idler photons and signal photons recalled from channel 2 (mode numbers: 1, 80, 160, 240, 330). The average $g_{s,i}^{(2)}(0)$ between correlated recalled signal and idler photon modes (1650 pairs) is 22.92 $\pm$ 0.07. For uncorrelated modes, the average $g_{s,i}^{(2)}(0)$ is 1.01 $\pm$ 0.01. Error bars in Figs. (b) and (d) are calculated by standard deviations of counts which obey Poisson distribution.}
\label{fig:3}
\end{figure}

Second, to examine the effect of crosstalk between different spectral channels, we measure the $g_{s,i}^{(2)}(0)$ between recalled signal photons and heralding idler photons corresponding to different spectral channels\cite{pu2017experimental}. If there is no crosstalk between different spectral channels, the measured $g_{s,i}^{(2)}(0)$ between uncorrelated channels should be around 1. On the other hand, with increasing crosstalk, the value of $g_{s,i}^{(2)}(0)$ would also increase (the relationship between crosstalk and $g_{s,i}^{(2)}(0)$ is shown in Supplementary Information Note 5). To eliminate the crosstalk between spectral channels, a sufficiently large separation between adjacent AFCs is required\cite{sinclair2014spectral} - in our case we choose 5 GHz separation. We first assess the crosstalk resulting from the source and detection system without the quantum memory. For the crosstalk characterization, we label the signal photons corresponding to channels 1 to 5 as $S_{1}$, $S_2$, ..., $S_5$, and similarly, the idler photons as $I_1$, $I_2$, ..., $I_5$. The choice of which signal and idler channels to detect is determined by the tunable FBG filter settings. Measuring coincidences between $S_m$ and $I_n$, where $m$, $n$ $\in$\{1, 2, 3, 4, 5\}, we calculate a 5 $\times$ 5 matrix of $g_{s,i}^{(2)}(0)$. The measured average $g_{s,i}^{(2)}(0)$ is 27.90 $\pm$ 0.18 for 5 pairs of correlated spectral modes (m=n) and 1.07 $\pm$ 0.02 for the 20 pairs of uncorrelated spectral modes (m$\ne$n), where error bars are acquired through Monte Carlo simulation. The detailed values are shown in Supplementary Information Note 6. The results indicate that the crosstalk between different spectral channels is negligible. Next, we evaluate whether the multimode memory induces additional crosstalk. To that end, we send the signal photons into the five AFC spectral channels with the storage time of 200 ns. The recalled signal photons from different channels are indexed as $R_1$, $R_2$, ..., $R_5$. Measuring coincidences between $R_m$ and $I_n$, we obtain the recalled signal photons from five channels (see Fig. \ref{fig:3}(a)). The transmitted photons output from the EDF indicate the optical delay of the fibre, and the time delay between the transmitted photons and the recalled photons shows the storage time of the AFC memory. We again calculate the 5 $\times$ 5 array of $g_{s,i}^{(2)}(0)$ values as shown in Fig. \ref{fig:3}(b). The measured average $g_{s,i}^{(2)}(0)$ between recalled photons and idler photons is 25.48 $\pm$ 1.05 for five correlated spectral modes and 1.01 $\pm$ 0.10 for 20 pairs of uncorrelated spectral modes. The results confirm that the crosstalk between different spectral channels is negligible in our quantum memory.

Third, to quantify the multimode capacity of our quantum memory, we create a train of heralded single photons - with pulse durations of and spaced by 300 ps - for simultaneous storage in all five spectral channels. With the storage time of 200 ns, 330 temporal modes are simultaneously stored in each spectral channel (see Fig. \ref{fig:3}(c)). The total stored mode number is, thus, up to 1650. Furthermore, to assess the crosstalk between different temporal modes, we obtain 1650 $\times$ 1650 values of $g_{s,i}^{(2)}(0)$ through measuring cross-correlation function between different idler photons and recalled signal photons. Figure \ref{fig:3}(d) presents parts of $g_{s,i}^{(2)}(0)$ corresponding to different temporal modes in channel 2 (the full 1650 $\times$ 1650 array of $g_{s,i}^{(2)}(0)$ is shown in Supplementary Information Note 7). The average $g_{s,i}^{(2)}(0)$ between correlated signal and idler photonic modes (1650 pairs) is 22.92 $\pm$ 0.07, verifying that the non-classical correlations are intact. For uncorrelated modes the $g_{s,i}^{(2)}(0)$ is 1.01 $\pm$ 0.01 on average with error bar from Monte Carlo simulation, confirming the negligible crosstalk between different temporal modes. In addition, we conclude the key metrics of our multimode quantum memory in Table II, which also includes the state-of-the-art of multimode quantum storage of non-classical light based on REID materials.
\section{\\Discussion}
\\An important feature of our demonstration is the large multimode capacity of quantum memory for storage of single photons at telecom wavelength. Yet there are clear avenues to further advance the multimode capacity. First, by making use of the entire THz-wide inhomogeneous broadening of the EDF, at least seventy individual 10-GHz-wide AFCs could be generated with a separation of 5 GHz. Generating more large-bandwidth channels requires a broader optical frequency comb, which can be achieved by a nonlinear broadened comb\cite{hu2018single}, Kerr soliton microcomb\cite{yao2018gate,geng2021coherent}, and mode-locked fibre laser comb\cite{hillerkuss201126} etc. Second, by increasing the temporal mode number in each channel to the upper bound determined by the time-bandwidth product of the AFC, a total number of spectro-temporal modes of 10 GHz $\times$ 230 ns = 2300 can be realized in each channel. Combining with these improvements, the multimode capacity would exceed 70 $\times$ 10 GHz $\times$ 230 ns $\approx$ 160000. In addition, we note that the multimode capacity of our quantum memory could be further increased via adapting multiplexing in the spatial degree\cite{lan2009multiplexed,grodecka2012high,nicolas2014quantum,ding2015quantum,zhou2015quantum,parniak2017wavevector,pu2017experimental,chrapkiewicz2017high,tian2017spatial,wang2020large}.

Several upgrades could be applied to develop quantum memories available for future quantum networks, including storage time and storage efficiency. The storage time of AFC quantum memory is determined by the teeth spacing ($\Delta$) of AFC, i.e., T$_{storage}$=1/$\Delta$. In our demonstration, the frequency stability of laser system for AFC preparation limits the $\Delta$ to several MHz, resulting in a storage time of hundreds of nanoseconds. A straightforward improvement is to employ a laser system with Hertz linewidth to prepare AFC. By doing so, the storage time would be further limited by the homogeneous linewidth of Er$^{3+}$ ions in fibre, i.e., inversely proportional to its optical coherence time. In our experiment, the optical coherence time reaches tens of microsecond with a temperature of 10 mK and a magnetic field of 0.2 T. The optical coherence property can be further improved, for instance by reducing the spin flip flops and superhyperfine broadening. The interaction of Er$^{3+}$-Er$^{3+}$ spin flip flops can be reduced by decreasing the doping concentration of Er$^{3+}$ ions\cite{veissier2016optical}. And the superhyperfine broadening due to the interaction between Er$^{3+}$ ions and neighboring Al$^{3+}$ ions could be modified by removing the co-dopants of EDF\cite{staudt2006investigations}. In addition, it is desirable to develop on-demand quantum memories. A promising option to enable this feature is spin-wave storage\cite{afzelius2010demonstration,ma2021elimination}, which allows optical coherence to be mapped onto a long-lived spin state, such as the hyperfine level. The $\Lambda$-like structure has been harnessed in erbium ions\cite{baldit2010identification}, which demonstrated erbium ions doped solids has potential for realizing on-demand quantum memory. With regards to the storage eﬀiciency, the non-zero background absorption (d$_0$) of the prepared AFC is one of main limits. In our demonstration, d$_0$ is caused by the limited ratio of the lifetime of Zeeman sublevels ($\sim$300 ms) to the waiting time following the AFC preparation (200 ms), in which the waiting time is to ensure that recalled photons are not masked by spontaneously emitted photons and guarantee the memory with high signal-to-noise ratio (SNR $>$ 1000). Along with the waiting time, the population of Zeeman sublevels decays accordingly, which inevitably fills the AFC troughs\cite{saglamyurek2015quantum,saglamyurek2016multiplexed}. To further increase the efficiency of the memory, on one hand, the lifetime of Zeeman sublevels could be increased by decreasing the doping concentration of Er$^{3+}$ ions\cite{saglamyurek2015quantum}. On the other hand, the trade-off between the waiting time and the SNR of memory should be explored, i.e., appropriately sacrificing the SNR of memory by reducing the waiting time to improve the storage efficiency. Furthermore, utilizing an impedance-matched optical cavity could also improve the storage efficiency\cite{afzelius2010impedance,moiseev2010efficient}. Inspired by these methods, towards a high-performance EDF based AFC quantum memory, a favorable avenue is to create an impedance-matched in-fibre cavity in a piece of EDF with low doping concentration (more details see Supplementary Information Note 8).

In summary, we have demonstrated a multimode quantum memory that is suitable for storage of spectro-temporal modes of single photons at telecom wavelength. A quantum memory with five 10-GHz-wide AFC channels in a 10-m-long cryogenically cooled EDF is prepared, and 330 temporal modes of heralded single photons are stored in each channel leading to the multimode capacity up to 1650. The key method introduced here is the combination of an optical frequency comb with frequency chirping to prepare large-bandwidth multi-channel AFCs, thus, enabling the large multimode capacity and high spectral isolation of our memory. With the improvements of storage time and efficiency, EDF based AFC quantum memory would pave the way for constructing future quantum internet compatible with current telecom infrastructure.

\section{\\Methods}\\
\textbf{Erbium doped silica fibre.} The experiment utilizes a 10-m-long, single-mode, commercial EDF with Er$^{3+}$ ions doping concentration of 200~ppm, co-doped with Al, Ge, and P. The fibre is spooled to a home-made copper cylinder with a diameter of 4~cm and fixed in the dilution refrigerator (LD400, Bluefors), in which the temperature can be cooled below 10 mK. The measured absorption at 1532 nm is 0.35~dB/m at T=10 mK. The EDF is fused with single-mode fibres for each end, and is exposed to 2000~Gauss magnetic field, the strength of which is optimized by observing the lifetime of Zeeman sublevels of $^4$I$_{15/2}$ level. With a magnetic field of 0.2 T, the lifetime reaches 0.278 $\pm$ 0.035 s, which ensures the persistent time of AFC (see Supplementary Information Note 9 for details). On the other hand, the applied magnetic field can eliminate a part of magnetic two-level systems, thus increasing the optical coherence time of erbium ions\cite{macfarlane2006optical}. The loss of the whole fibre sample is about 2.5 dB, including bending, splicing, and transmission loss.
\\
\textbf{Experimental time sequences and procedures.} The time sequences are shown in Fig. \ref{fig:2}(b). During the preparation time of 300 ms, the pump light from a CW-laser operating at 1531.88 nm is prepared into an optical frequency comb with frequency spacing of 15 GHz by an IM and a PM, both of which are driven by 15 GHz microwave signals. The optical frequency comb is sent to another PM, which is driven by an arbitrary waveform generator (AWG) to continuously generate frequency chirping light on each comb ranging from -5 GHz to 5 GHz (total span of 10 GHz). Then the prepared pump light is modulated into 300-ms-long light pulses by an OS, and sent into the EDF to prepare quantum memory with five spectral channels. The setting 200 ms waiting time is to decrease the influence of spontaneously emitted photons from excited ions on the recalled photons from the AFC (see Supplementary Information Note 10 for more details). During the storage time of 500 ms, another OS turns on. The light from a CW-laser operating at 1540.60 nm is modulated into a light pulse with pulse duration of 300 ps and the repetition rate of 1 MHz (for multiple temporal modes storage, the laser is modulated to 330 light pulses within one cycle of 1 $\mu$s, i.e., T$_{period}$=1 $\mu$s) and sent to PPLN module for the generation of correlated photon-pairs with a bandwidth of $\sim$60~nm. By utilizing DWDMs, signal photons at 1531.88~nm and idler photons at 1549.32~nm with a bandwidth of 100~GHz are filtered out. The signal photons are sent into five-channel quantum memory to filter and to store. After the storage time of T$_{storage}$, the signal photons are recalled from AFCs. Finally, the idler photons and recalled signal photons corresponding to different channels are selected by FBGs and detected by SNSPDs, subsequently performed coincidence analysis in the TDC.
\\
\textbf{Superconducting nanowire single photon detectors.}  All detections of single photons are carried out by a set of SNSPDs system provided by PHOTEC Corp. The system consists of SNSPD devices, cryostat system, and electronic control system. The niobium nitride (NbN) SNSPDs, manufactured by Shanghai Institute of Microsystem and Information Technology (SIMIT), operate at $\sim$2.2 K in the cryostat system with a dark counting rate of $\sim$100 Hz and a time jitter of $\sim$100 ps. All detectors have a dead time of less than 50 ns and a detection efficiency of $\sim$60\%.
\\
\textbf{Calculation of cross-correlation function $g_{s,i}^{(2)}(0)$.} Considering the counting effective events in $m$ experimental trials, we record the total 3-fold coincidence counts $C_{si}$ of the trigger signal, idler and signal photons, the coincidence counts $C_s$ ($C_i$) between signal (idler) photons and trigger signals. With the photon detected probability in each trial calculated as $p_{si}=C_{si}/m$, $p_s=C_s/m$, $p_i=C_i/m$, the second-order cross-correlation function $g_{s,i}^{(2)}(0)$ is calculated by
\begin{equation}
g_{s,i}^{(2)}(0)=\frac{p_{si}}{p_s\cdot p_i}=\frac{C_{si}\cdot m}{C_s \cdot C_i}.
\end{equation}

It is a strong indication of quantum correlations for photons-pairs if $g_{s,i}^{(2)}(0)\gg2$. According to Cauchy-Schwarz inequality, it also indicates that the auto-correlation function of heralded signal photons is $\ll$1, i.e., these signal photons are denoted as single photons\cite{bashkansky2014significance}. For all the coincidence measurements in this paper, the width of coincidence window is 600 ps.
\\
\section{\\Data availability}\\
The data that support the findings of this study are available from the corresponding author on reasonable request.

\section{\\Acknowledgments}\\
This work was supported by the National Key Research and Development Program of
China (Nos. 2018YFA0307400, 2018YFA0306102), National Natural Science Foundation of China (Nos. 61775025, 91836102, U19A2076, 12004068), Innovation Program for Quantum Science and Technology (No. 2021ZD0301702), Sichuan Science and Technology Program (Nos. 2021YFSY0062, 2021YFSY0063, 2021YFSY0064, 2021YFSY0065, 2021YFSY0066), China Postdoctoral Science Foundation (Nos. 2020M683275, 2021T140093).

\section{\\Author contributions}\\
 D.O., G.G., and Q.Z. conceived and supervised the project. S.W. and B.J. mainly carried out the experiment and collected the experimental data with help of other authors. H.L., L.Y. and Z.W. developed and maintained the SNSPDs used in the experiment. S.W., B.J. and Q.Z. analyzed the data. S.W., B.J., D.O. and Q.Z. wrote the manuscript with inputs from all other authors. All authors have given approval for the final version of the manuscript.

\section{\\Competing interests}\\
\\The authors declare no competing interests.

\bibliography{myref}

\clearpage

\clearpage

\clearpage

\clearpage
\begin{table}[htbp]\color{black}
	\caption{Measured crosstalk ($g_{s,i}^{(2)}(0)$) between recalled signal photons and idler photons with different storage time.}
	\label{table2}
	\begin{ruledtabular}
		\begin{tabular}{c@{}c@{}c@{}c@{}c@{}c@{}}
		Storage time & Channel 1 & Channel 2 & Channel 3 & Channel 4 & Channel 5 \\ 
		\hline
    0 ns & 25.64$\pm$0.19 & 27.53$\pm$0.18 & 27.91$\pm$0.18 & 28.38$\pm$0.18 & 28.62$\pm$0.19 \\
   10 ns & 26.13$\pm$1.45 & 28.97$\pm$1.20 & 26.81$\pm$1.49 & 26.10$\pm$1.05 & 27.07$\pm$1.15 \\
   50 ns & 26.95$\pm$1.48 & 26.72$\pm$1.32 & 26.85$\pm$1.59 & 26.95$\pm$1.23 & 22.99$\pm$1.11 \\
  100 ns & 26.08$\pm$1.90 & 25.77$\pm$1.60 & 24.48$\pm$1.92 & 27.01$\pm$1.69 & 25.85$\pm$1.36 \\
  230 ns & 26.22$\pm$4.32 & 31.00$\pm$4.41 & 26.22$\pm$4.32 & 26.59$\pm$3.81 & 22.09$\pm$3.34 \\
		\end{tabular}
	\end{ruledtabular}
\end{table}

\clearpage
\begin{table}[htbp]\color{black}
	\caption{Comparison on the time-bandwidth product and multimode capacity of quantum memories for non-classical light based on different REID materials.}
	\label{table3}
	\begin{ruledtabular}
	\begin{tabular}{c@{}c@{}c@{}c@{}c@{}c@{}}
	System & $\lambda$\footnote{$\lambda$ is the wavelength of the stored single photons.} (nm) & T$_{storage}$\footnote{T$_{storage}$ is the storage time of AFC quantum memory.} (ns) & BW\footnote{BW is the available storage bandwidth of quantum memory.} (GHz) & TBP\footnote{TBP is the time-bandwidth product of quantum memory.} & N\footnote{Number of stored modes (i.e., multimode capacity).} \\ 
	\hline
    Eu$^{3+}$:Y$_{2}$SiO$_{5}$\cite{laplane2017multimode} & 580 & 1$\times$10$^6$ & 0.002 &2000 & 12 \\
	Pr$^{3+}$:Y$_{2}$SiO$_{5}$\cite{seri2019quantum} & 606 & 3500 & 0.06 &210 & 130 \\
	Tm$^{3+}$:Ti$^{3+}$:LiNbO$_{3}$\cite{saglamyurek2011broadband} & 795 & 7 & 5 & 35 & 1 \\
	Tm$^{3+}$:Y$_{3}$Al$_{5}$O$_{12}$\cite{davidson2020improved} & 795 & 100 & 0.5 & 50 &1 \\
	Nd$^{3+}$:YVO$_{4}$\cite{tang2015storage} & 880 & 500 & 0.5 & 250 & 100 \\
	Nd$^{3+}$:Y$_{2}$SiO$_{5}$\cite{clausen2011quantum,tiranov2016temporal} & 883 & 200 & 0.12 & 24 & 10 \\
	Yb$^{3+}$:Y$_{2}$SiO$_{5}$\cite{businger2022non} & 979 & 2.5$\times$10$^4$ & 0.1 & 2500 & 1250 \\
	Er$^{3+}$:Ti$^{3+}$:LiNbO$_{3}$\cite{askarani2019storage} & 1532 & 48 & 6 & 288 & 1 \\
	Er$^{3+}$:LiNbO$_{3}$\cite{Zhang2022storage} & 1532 & 100 & 4 & 400 & 147 \\
	EDF\cite{saglamyurek2016multiplexed} & 1532 & 50 & 16 & 800 & 6 \\
	\textbf{EDF (this work)} & \textbf{1532} & \textbf{230} & \textbf{50} & \textbf{11500} & \textbf{1650} \\
	\end{tabular}
	\end{ruledtabular}
\end{table}
\clearpage

\section{\centerline{Supplementary Information}}
\renewcommand{\baselinestretch}{1.5}
\renewcommand\figurename{Fig.S.}
\renewcommand\tablename{TABLE S.}
\renewcommand{\andname}{\ignorespaces}
\setcounter{figure}{0}
\setcounter{table}{0}
\subsection*{Note 1: Preparation of quantum memory with five spectral channels}
\begin{figure}[ht]
\centering
\includegraphics[]{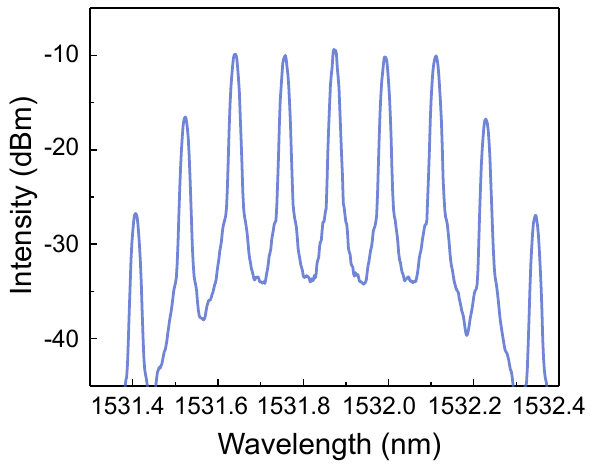}
\caption{\textbf{Optical frequency comb.} The central wavelengths of five combs with same intensity are 1532.11 nm, 1532.00 nm, 1531.88 nm, 1531.76 nm, 1531.65 nm, respectively.}
\label{fig:1-1}
\end{figure}

\noindent \\Due to the inhomogeneous broadening in erbium doped silica fibre (EDF) up to THz, EDF is usually utilized to prepare broadband quantum memory\cite{saglamyurek2015quantum,askarani2020entanglement,saglamyurek2016multiplexed} and can in principle support the storage of ps-width photonic wavepackets. In this section, we present the preparation of AFCs with five spectral channels in EDF. First, the light from a continuous-wave laser (CW-laser) operating at 1531.88 nm is modulated into an optical frequency comb by cascading a phase modulator (PM) and an intensity modulator (IM)\cite{wu2010generation}, at which IM is used to improve spectral flatness of the generated optical frequency comb\cite{dou2011improvement}. The 15~GHz microwave signal is divided into two parts. One is sent directly to the PM, and the other is sent to the IM with a phase shifter connected, resulting in an optical frequency comb with a frequency spacing of 15~GHz, as shown in Fig.S. \ref{fig:1-1}. 
The prepared optical frequency comb is sent to another PM to generate frequency chirping light on each comb. The modulated signal is composed of a series of sine waves with frequencies from 0 to 5~GHz. The $\pm$1st order frequency-shifted light generated with PM has the same intensity. Therefore, by adding 0 -- 5~GHz modulation signals to the PM, the frequency shift of the light ranges from -5~GHz to 5~GHz. In this way, we prepare five AFCs in the EDF with frequency chirping 10 GHz bandwidth for each channel.

\begin{figure}[ht]
\centering
\includegraphics[]{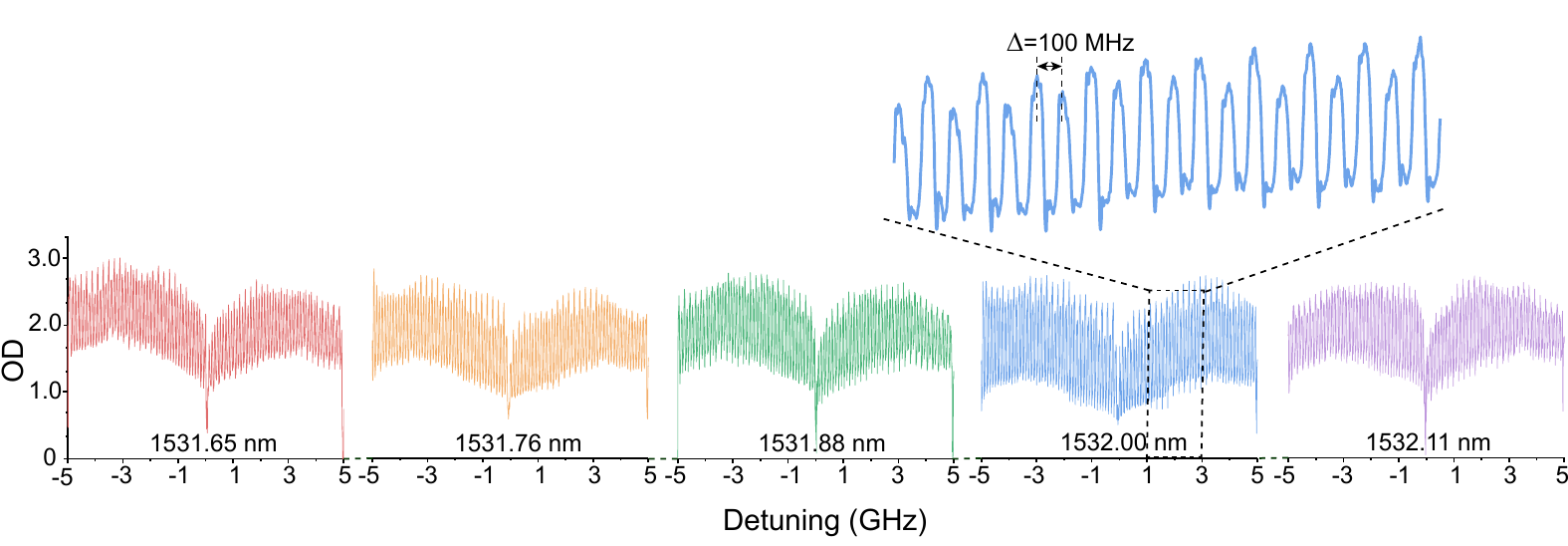}
\caption{\textbf{Typical AFCs with five spectral channels.} A typical example of AFCs with five spectral channels prepared by utilizing optical frequency comb and frequency chirping method. The bandwidth of each AFC is 10 GHz, and the spacing of adjacent AFCs is 5 GHz. In total, the available bandwidth for quantum memory is 50 GHz. Inset: Details for a 2-GHz-wide section of the AFC is shown, where the comb spacing $\Delta$ is 100 MHz. OD: Optical Depth.}
\label{fig:1-2}
\end{figure}

In our experiment, we prepare five AFCs at the central wavelengths of 1532.11 nm, 1532.00 nm, 1531.88 nm, 1531.76 nm and 1531.65 nm, respectively. A typical example of five AFCs is shown in Fig.S. \ref{fig:1-2}.

\subsection*{Note 2: Heralded single photon source}
\begin{figure}[ht]
\centering
\includegraphics[width=14cm]{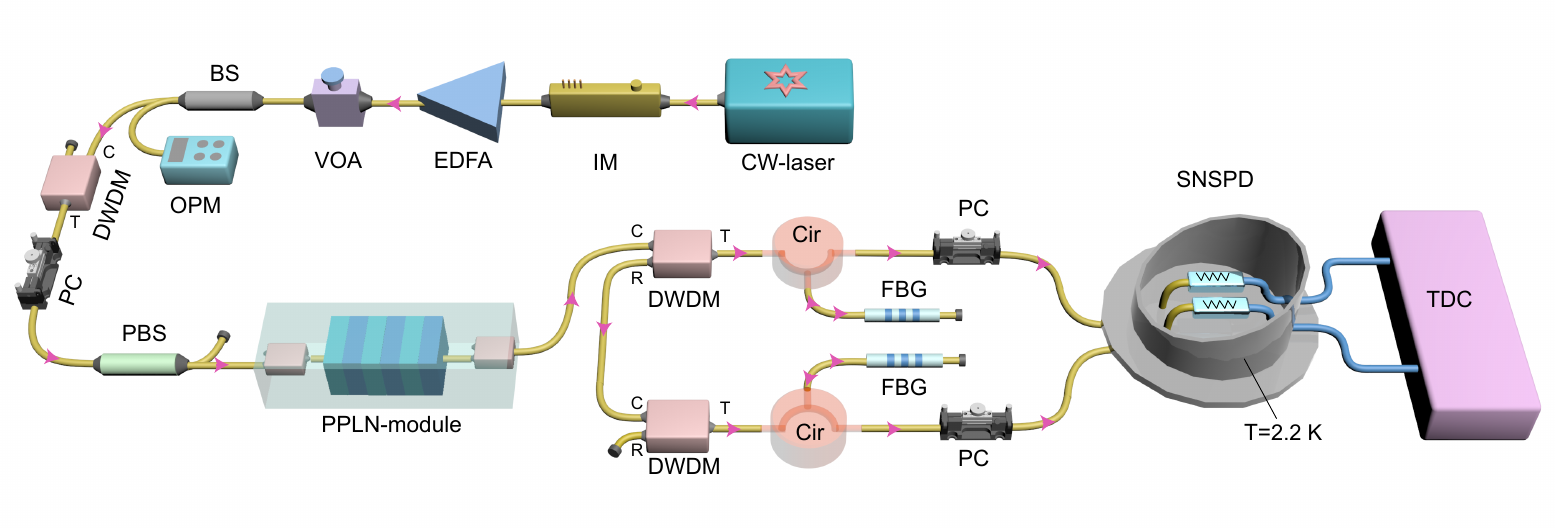}
\caption{\textbf{Experimental setup for heralded single photon source at telecom wavelength.} CW-laser: continuous-wave laser, IM: intensity modulator, EDFA: erbium doped fibre amplifier, VOA: variable optical attenuator, BS: beam splitter, OPM: optical power meter, PC: polarization controller, PBS: polarizing beam splitter, PPLN-module: periodically poled lithium niobate module, DWDM: dense wavelength division multiplexer, Cir: circulator, FBG: fibre Bragg grating, SNSPD: superconducting nanowire single photon detector, TDC: time-to-digital converter.}
\label{fig:2-1}
\end{figure}

\noindent \\In this section, we present properties of the heralded single photon source at telecom wavelength corresponding to different AFC spectral channels. The experimental setup is shown in Fig.S. \ref{fig:2-1}. The scheme for generating correlated photon-pairs is cascaded second-harmonic generation (SHG) and spontaneous parametric down conversion (SPDC) processes in a single periodically poled lithium niobate (PPLN) waveguide\cite{arahira2012experimental,arahira2011generation,hunault2010generation,zhang2021high,yu2022spectrally} (the main parameters of the PPLN see Table S. I).
\begin{table}[ht]\color{black}
\centering
	\caption{Main parameters of PPLN module.}
	\label{table1}
\begin{tabular}{l l c}
\hline
\hline
 Type of waveguide & & RPE waveguide \\
 Length of waveguide & & 50 mm \\
 Poled period & & 19 $\mu$m \\
 SHG normalized conversion efficiency & & 336.07\%/W\\
 Length of pigtail & & 20 cm \\
 Input coupling efficiency of PPLN waveguide & & 55.51\% \\
 Output coupling efficiency of PPLN waveguide & & 77.89\% \\
  \hline
  \hline
\end{tabular}
\end{table}
The light from a narrow linewidth continuous-wave laser operating at 1540.60 nm is modulated into light pulses with a repetition rate of 1 MHz by an intensity modulator (IM) and sent into the PPLN module, resulting in correlated photon-pairs with a bandwidth of $\sim$60 nm. Erbium doped fibre amplifier (EDFA) and variable optical attenuator (VOA) are utilized to adjust the optical power of pump light, and polarization controller (PC) and polarizing beam splitter (PBS) are used to adjust the polarization of pump light. Dense wavelength division multiplexers (DWDMs) are utilized to select signal photons (1531.88 nm) and idler photons (1549.32 nm) with a bandwidth of 100~GHz. The generated correlated photon-pairs are then sent directly to the detection system. In the measurement system, correlated photon-pairs with a bandwidth of $\sim$6 GHz corresponding to five channels of quantum memory are filtered out by adjusting the temperatures of two tunable fibre Bragg gratings (FBGs). The central wavelengths of signal photons corresponding to channels 1 to 5 are 1532.11 nm, 1532.00 nm, 1531.88 nm, 1531.76 nm, and 1531.65 nm, respectively. Correspondingly, the central wavelengths of idler photons are 1549.08 nm, 1549.20 nm, 1549.32 nm, 1549.44 nm, and 1549.56 nm. Finally, the selected correlated photon-pairs are detected by two superconducting nanowire single photon detectors (SNSPDs). Circulators (Cirs) are utilized to collect the reflected photons from FBGs. 

We measure the photons counting rate of correlated photon-pairs corresponding to different spectral channels under different power of pump light.
\begin{figure}[ht]
\centering
\includegraphics[]{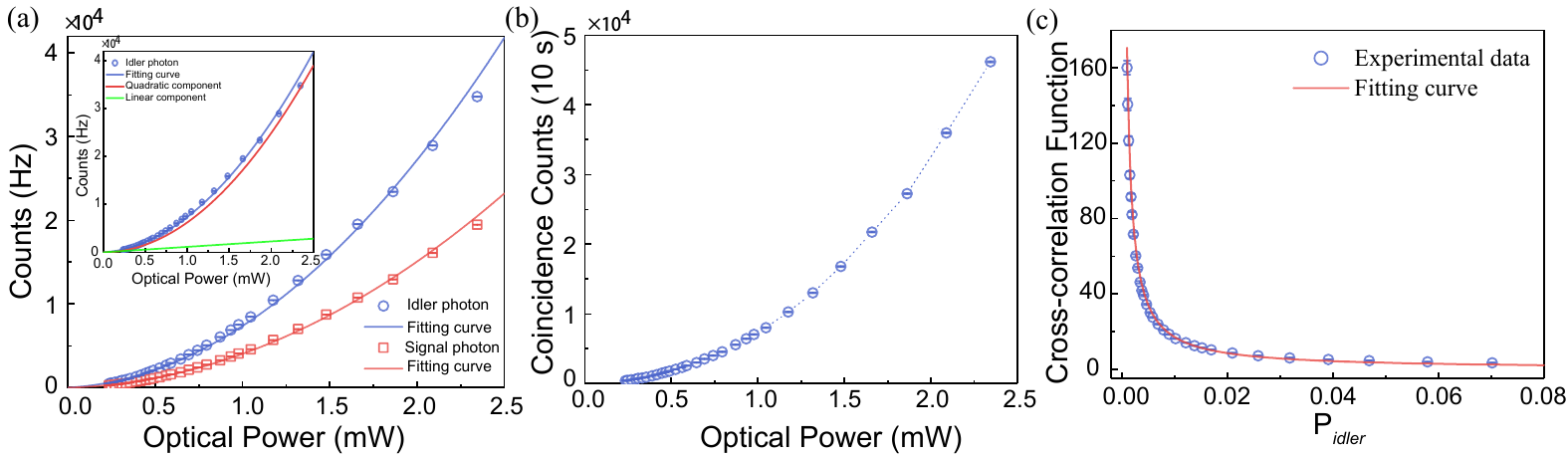}
\caption{\textbf{Measured results of correlated photon-pairs related to channel 2.} (a) Photons counting rate of signal photons and idler photons versus pump power. Inset: Idler photon-counting rate (blue circle) versus pump power. The blue solid line is the quadratic polynomial fitting curve of the photon-counting rate with the quadratic and linear parts shown as
the red and green curves, respectively. The quadratic and linear components  are corresponding to the contributions of generated photon-pairs and noise photons. (b) Coincidences counting rate between signal photons and idler photons versus pump power. (c) Cross-correlation function ($g_{s,i}^{(2)}(0)$) of correlated photon-pairs versus P$_{idler}$. All error bars are calculated by standard deviations of counts which obey Poisson distribution.}
\label{fig:2-2}
\end{figure}
Fig.S. \ref{fig:2-2}(a) shows the photons counting rate of a typical correlated photon-pairs corresponding to channel 2 with different pump powers. The blue circles and red squares represent idler photons and signal photons, respectively. Here, the counting rates of signal photons and idler photons fit well with the quadratic polynomial curves. The measured counting rate of idler photons is slightly higher than signal photons. This is mainly owning to the difference in transmission efficiency and bandwidth of the filters. The total transmission loss of idler photons is 3.9 dB, including the loss of DWDM, PC, FBG, etc., while the loss of signal photons is 5.5 dB. The bandwidth of the FBG with the central wavelength of 1549.32 nm (1531.88 nm) is $\sim$0.06 nm (0.05 nm). Fig.S. \ref{fig:2-2}(b) shows the coincidence counting of idler photons and signal photons with different pump powers. Then we measure the cross-correlation function ($g_{s,i}^{(2)}(0)$) of photon-pairs corresponding to different channels of quantum memory under different generation probabilities of idler photons. Fig.S. \ref{fig:2-2}(c) shows $g_{s,i}^{(2)}(0)$ for typical correlated photon-pairs corresponding to channel 2 of the quantum memory with different generation probabilities of idler photons (P$_{idler}$). The measured $g_{s,i}^{(2)}(0)$ fits well with an inversely proportional curve. Considering that the output coupling efficiency of the PPLN module is 71.65\%, the total transmission efficiency is 40.74\%, and the detection efficiency of SNSPD is 60\%, the measured efficiency of idler photons is 17.51\%. For a measured P$_{idler}$ of 0.62\%, the intrinsic generation probability of idler photons in the waveguide is 3.54\%, which results in a measured $g_{s,i}^{(2)}(0)$ of $27.53\pm0.18$. Detailed values corresponding to each channel are shown in Table S. II. 

\begin{table}[ht]\color{black}
	\caption{Values of $g_{s,i}^{(2)}(0)$ for correlated photon-pairs corresponding to different channels.}
	\label{table2}
	\begin{ruledtabular}
		\begin{tabular}{c@{}c@{}c@{}c@{}c@{}c@{}}
		Channel\footnote{Correlated photon-pairs corresponding to different channels of quantum memory.} & Signal photon (nm)\footnote{Central wavelengths of signal photons corresponding to different channels.} & Idler photon (nm)\footnote{Central wavelengths of idler photons corresponding to different channels.} & $g_{s,i}^{(2)}(0)$ \\ \hline
			1	& 1532.11 & 1549.08 & 25.64 $\pm$ 0.19   \\	
			2	& 1532.00 & 1549.20 & 27.53 $\pm$ 0.18   \\
			3	& 1531.88 & 1549.32 & 27.91 $\pm$ 0.18   \\
			4	& 1531.76 & 1549.44 & 28.38 $\pm$ 0.18  \\
			5	& 1531.65 & 1549.56 & 28.62 $\pm$ 0.19 \\
		\end{tabular}
	\end{ruledtabular}
\end{table}

\subsection*{Note 3: Performance of the five-channel quantum memories}
\begin{figure}[htbp]
\centering
\includegraphics[]{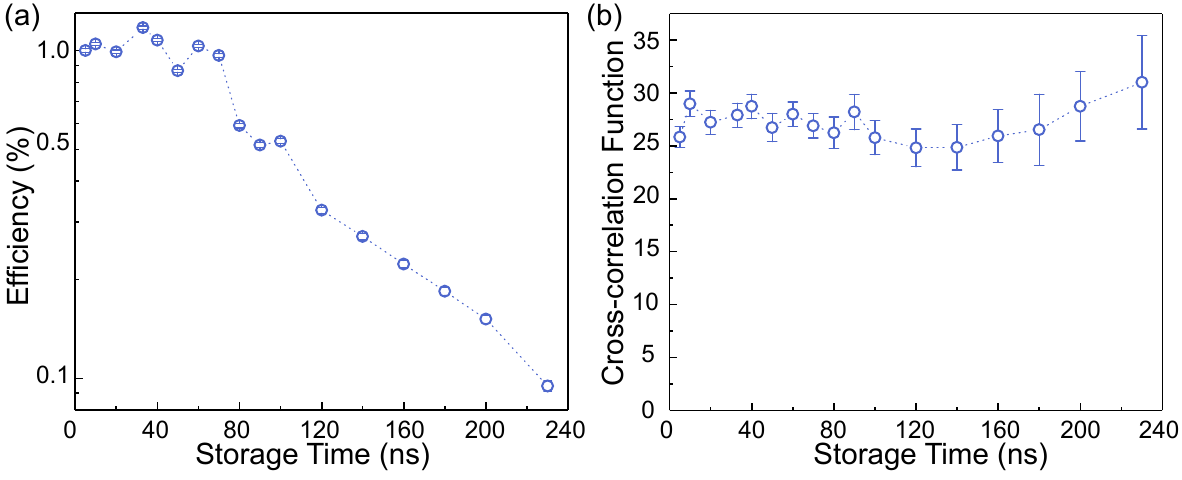}
\caption{\textbf{Performance of multi-channel quantum memory.} (a) A typical example (channel 2) for system storage efficiency as a function of storage time. The comb tooth spacing are prepared ranging from 200 to 4.35 MHz, which determines the storage time from 5 to 230 ns. (b) A typical example (channel 2) for cross-correlation function after storage as a function of storage time. The measurement time for each point is 1000~s. All error bars are calculated by standard deviations of counts which obey Poisson distribution.}
\label{fig:3-1}
\end{figure}

\noindent \\In this section, we first take channel 2 as an example to illustrate the storage performance. Fig.S. \ref{fig:3-1}(a) shows the system storage eﬀiciency, defined as the ratio of the count rate of recalled heralded photons to that of the input heralded photons, for a typical channel of quantum memory (channel 2) versus storage time. At small storage time ($<$ 100 ns), a slight oscillation of the storage efficiency is visible. This may be due to a quantum beat generated by the interaction of the Er$^{3+}$ with other surrounding ions (such as the co-doped Al$^{3+}$)\cite{de2008solid,staudt2006investigations}. For longer storage time, the decay is exponential with decay constant of 48 ns. We attribute this to the decoherence effects of Er$^{3+}$ and the imperfect preparation of AFCs\cite{saglamyurek2015quantum}. Fig.S. \ref{fig:3-1}(b) shows $g_{s,i}^{(2)}(0)$ after storage as a function of storage time for channel 2. Unlike the efficiency, the cross-correlation function $g_{s,i}^{(2)}(0)$ remains almost constant at a level above 25 even at the maximum storage time of 230 ns. First and foremost, this shows that non-classical properties of the correlated photon-pairs are preserved in the memory and, in addition, that even at the largest storage time the signal-to-noise ratio (SNR) of the system is high enough ($>$ 730) that the noise does not degrade the cross-correlation function. (see Note 4 for more details on the role of SNR in determining $g_{s,i}^{(2)}(0)$ in our quantum memory). The cross-correlation function after storage versus storage time for all five-channel quantum memories are shown in Table S. III.

\begin{table}[htbp]\color{black}
	\caption{Performance of each channel of quantum memory.}
	\label{table3}
	\begin{ruledtabular}
		\begin{tabular}{c@{}c@{}c@{}c@{}c@{}c@{}}
		Storage time & Channel 1 & Channel 2 & Channel 3 & Channel 4 & Channel 5 \\ 
		\hline
   0 ns & 25.64$\pm$0.19 & 27.53$\pm$0.18 & 27.91$\pm$0.18 & 28.38$\pm$0.18 & 28.62$\pm$0.19 \\
   5 ns & 25.89$\pm$1.38 & 25.82$\pm$0.97 & 26.69$\pm$1.49 & 26.10$\pm$1.06 & 25.23$\pm$1.12 \\	
   10 ns & 26.13$\pm$1.45 & 28.97$\pm$1.20 & 26.81$\pm$1.49 & 26.10$\pm$1.05 & 27.07$\pm$1.15 \\
   20 ns & 25.97$\pm$1.49 & 27.23$\pm$1.13 & 26.48$\pm$1.50 & 27.35$\pm$1.05 & 25.26$\pm$1.15 \\
   33 ns & 26.10$\pm$1.37 & 27.90$\pm$1.13 & 25.38$\pm$1.31 & 27.12$\pm$1.11 & 25.26$\pm$0.93 \\
   40 ns & 26.13$\pm$1.37 & 28.73$\pm$1.16 & 26.32$\pm$1.26 & 28.36$\pm$1.46 & 25.29$\pm$0.92 \\
   50 ns & 26.95$\pm$1.48 & 26.72$\pm$1.32 & 26.85$\pm$1.59 & 26.95$\pm$1.23 & 22.99$\pm$1.11 \\
   60 ns & 27.05$\pm$1.53 & 27.99$\pm$1.15 & 26.95$\pm$1.54 & 26.01$\pm$1.00 & 25.49$\pm$0.91 \\
   70 ns & 25.01$\pm$1.54 & 26.89$\pm$1.14 & 25.02$\pm$1.43 & 25.39$\pm$1.25 & 26.08$\pm$1.03 \\
   80 ns & 25.64$\pm$1.78 & 26.24$\pm$1.48 & 26.21$\pm$1.81 & 25.54$\pm$1.44 & 24.64$\pm$1.19 \\
   90 ns & 25.65$\pm$1.80 & 28.19$\pm$1.67 & 25.65$\pm$1.79 & 27.35$\pm$1.57 & 24.97$\pm$1.28 \\
  100 ns & 26.08$\pm$1.90 & 25.77$\pm$1.60 & 24.48$\pm$1.92 & 27.01$\pm$1.69 & 25.85$\pm$1.36 \\
  120 ns & 27.12$\pm$2.31 & 24.80$\pm$1.78 & 27.05$\pm$2.33 & 28.70$\pm$1.69 & 22.95$\pm$1.50 \\
  140 ns & 25.00$\pm$2.31 & 24.85$\pm$2.15 & 23.97$\pm$2.36 & 24.46$\pm$2.25 & 27.60$\pm$1.93 \\
  160 ns & 23.99$\pm$2.76 & 25.93$\pm$2.43 & 22.36$\pm$2.63 & 28.26$\pm$2.78 & 23.71$\pm$1.78 \\
  180 ns & 24.99$\pm$3.25 & 26.53$\pm$3.35 & 24.89$\pm$3.51 & 26.53$\pm$3.25 & 23.45$\pm$2.34 \\
  200 ns & 25.46$\pm$3.17 & 28.75$\pm$3.37 & 20.76$\pm$3.27 & 26.05$\pm$3.17 & 25.57$\pm$2.60 \\
  230 ns & 26.22$\pm$4.32 & 31.00$\pm$4.41 & 26.22$\pm$4.32 & 26.59$\pm$3.81 & 22.09$\pm$3.34 \\
		\end{tabular}
	\end{ruledtabular}
\end{table}

\subsection*{Note 4: Contribution of SNR to $g_{s,i}^{(2)}(0)$}
\begin{figure}[ht]
\centering
\includegraphics[width=12 cm]{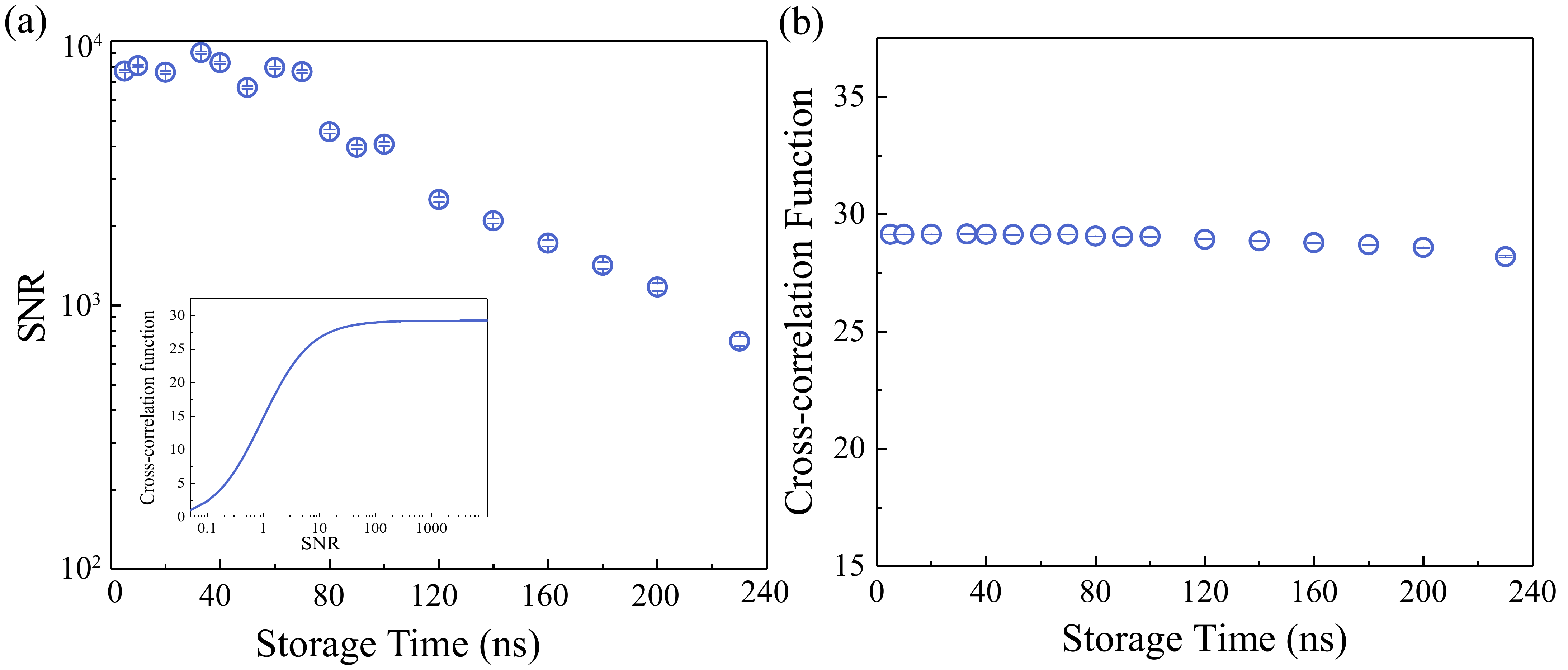}
\caption{\textbf{Cross-correlation function with different SNR.} (a) The SNR of quantum memory versus storage time. Inset: The estimated $g_{s,i}^{(2)}(0)$ versus SNR of quantum memory. (b) The results for the calculation of $g_{s,i}^{(2)}(0)$ with different storage time of quantum memory.}

\label{fig:4-1}
\end{figure}

\noindent \\This section we present the influence of SNR of quantum memory on quantum correlation of correlated photon-pairs, where the SNR is defined as the ratio of coincidence of idler-recalled photons to idler-noise photons. The measured 3-fold coincidence detecting probability $p_{si}$ of trigger signal, signal and idler photons, and the 2-fold coincidence detecting probability $p_s$ ($p_i$) of trigger signal and signal (idler) photons are expressed as
 \begin{equation}
  \begin{aligned}
     &p_{s}=\eta_c\eta_s+\eta_n,\\
   &p_{i}=\eta_c\eta_i+\eta_n,\\
  & p_{si}=\eta_c\eta_s\eta_i+p_{s}p_{i},
  \end{aligned}
\end{equation}
where $\eta_{c}$ is the internal generation probability of correlated photon-pairs in the waveguide, $\eta_{s}$ is the overall measured efficiency of recalled signal photons including transmission, storage, collection and detection efficiency, $\eta_{i}$ is the overall measured efficiency of idler photons, and $\eta_{n}$ is the detected probability of noise photons in each experimental trial, respectively. The $p_s\cdot p_i$ gives a good estimation for the accidental coincidences of uncorrelated background\cite{de2006direct}. Combining Eq. (1) with Eq. (2) in main text, $g_{s,i}^{(2)}(0)$ can be calculated as 
\begin{equation}
g_{s,i}^{(2)}(0)=\frac{p_{si}}{p_s\cdot p_i}=\frac{\eta_{c}\eta_{s}\eta_{i}}{(\eta_{c}\eta_{s}+\eta_{n})\cdot(\eta_{c}\eta_{i}+\eta_{n})}+1.
\end{equation}
During the experiment, $\eta_s$ is the only considered efficiency that will decrease with increasing storage time, and detection of idler and noise photons satisfies $\eta_c\eta_i\gg \eta_n$. The SNR of quantum memory system is defined as SNR$=\eta_{c}\eta_{i}\eta_{s}/(\eta_{c}\eta_{i}\eta_{n})=\eta_s/\eta_n$, where $\eta_{c}\eta_{i}\eta_{n}$ means the detected probability of coincidences between idler photons and noise photons. To analyze the contribution of SNR to $g_{s,i}^{(2)}(0)$, SNR is substituted into Eq. (2), then the $g_{s,i}^{(2)}(0)$ can be re-expressed as
\begin{equation}
g_{s,i}^{(2)}\simeq\frac{\eta_s}{\eta_c\eta_s+\eta_n}+1=\frac{1}{\eta_c+1/SNR}+1.
\end{equation}
If SNR is large enough, $g_{s,i}^{(2)}(0)$=$1/\eta_c+1$ will be independent with $\eta_s$, which means $g_{s,i}^{(2)}(0)$ will be unaffected by decreasing system storage efficiency once the overall measured efficiency ($\eta_s$) of recalled photons is still much larger than that of noise photons ($\eta_n$). \\
\indent
Considering the real experimental parameters, the internal generation probability of correlated photon-pairs is 3.54\%, and the measured coincidence probability of noise photons and idler photons (i.e., $\eta_{c}\eta_{i}\eta_{n}$) is $0.86\times10^{-9}$ in a 600-ps time window. According to the measured storage efficiency with different storage time, we calculate the SNR with different storage time, as shown in Fig.S. \ref{fig:4-1}(a). The SNR is 730 with storage time up to 230 ns. Besides, we calculate the $g_{s,i}^{(2)}(0)$ with different SNR as shown in the inset of Fig.S. \ref{fig:4-1}(a). It indicates that once the SNR is larger than 20, the $g_{s,i}^{(2)}(0)$ is nearly unaffected. It explains why $g_{s,i}^{(2)}(0)$ keeps unchanged  even with longer storage time and decreasing efficiency. According to the measured efficiency, we estimate the $g_{s,i}^{(2)}(0)$ with different storage time, as shown in Fig.S. \ref{fig:4-1}(b), which matches well with the experimental results shown in Note 3.

\subsection*{Note 5: Relationship of crosstalk and cross-correlation function}
\begin{figure}[ht]
\centering
\includegraphics[]{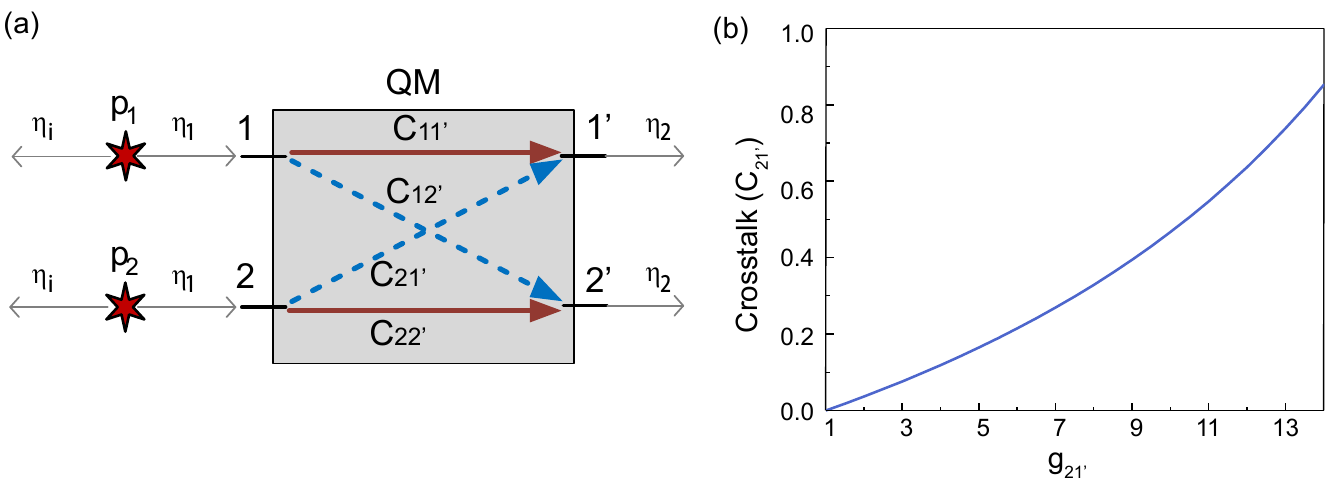}
\caption{\textbf{Relationship of crosstalk and cross-correlation function.} (a) A typical two-channel quantum memory with storage of two pairs of correlated photon-pairs. (b) Crosstalk between two different channels versus cross-correlation function of different channels.}
\label{fig:5-2}
\end{figure}

\noindent \\In this section, we describe the relationship of the crosstalk between different channels and the corresponding cross-correlation function $g_{s,i}^{(2)}(0)$. For simplicity, we consider a two-channel quantum memory, shown in Fig.S. \ref{fig:5-2}(a), where $1$ and $2$ represent the input channels of the quantum memory, $1'$ and $2'$ represent the corresponding outputs of the quantum memory, respectively.

Here we simply consider the crosstalk as the leakage efficiency of photons into the mismatched channels. The crosstalk coefficients of different channels can be expressed as $C_{11'}$, $C_{12'}$, $C_{21'}$, $C_{22'}$ with $C_{ij'}$ $(i,j=1,2)$ representing the efficiency of photons from channel $i$ passing through channel $j$ and satisfying $C_{11'}+C_{12'}=1$, $C_{12'}+C_{22'}=1$.

As shown in Fig.S. \ref{fig:5-2}(a), here we consider the situation of our experiment. $\eta_i$ is the overall measured efficiency of idler photons, $\eta_1$ is the transmission and collection efficiency of signal photons before quantum memory and $\eta_2$ is the measured efficiency of signal photons after the quantum memory, $p_1$ and $p_2$ are intrinsic generation efficiency of correlated photon-pairs. According to Eq. (2), the cross-correlation function corresponding to different channels can be expressed as
\begin{equation}
g_{11'}=\frac{p_{1}\eta_{i}\eta_{1}\eta_{2} C_{11'}+p_{1}\eta_{i}(p_1\eta_{1}\eta_{2} C_{11'}+p_{2}\eta_{1}\eta_{2}C_{21'})}{p_{1}\eta_{i}(p_1\eta_{1}\eta_{2} C_{11'}+p_{2}\eta_{1}\eta_{2}C_{21'})}=\frac{C_{11'}}{p_1C_{11'}+p_2C_{21'}}+1.
\end{equation}
\begin{equation}
g_{21'}=\frac{p_{2}\eta_{i}\eta_{1}\eta_{2} C_{21'}+p_{2}\eta_{i}(p_1\eta_{1}\eta_{2} C_{11'}+p_{2}\eta_{1}\eta_{2}C_{21'})}{p_{2}\eta_{i}(p_1\eta_{1}\eta_{2} C_{11'}+p_{2}\eta_{1}\eta_{2}C_{21'})}=\frac{C_{21'}}{p_1C_{11'}+p_2C_{21'}}+1.
\end{equation}
\begin{equation}
g_{12'}=\frac{p_{1}\eta_{i}\eta_{1}\eta_{2} C_{12'}+p_{1}\eta_{i}(p_1\eta_{1}\eta_{2} C_{12'}+p_{2}\eta_{1}\eta_{2}C_{22'})}{p_{1}\eta_{i}(p_1\eta_{1}\eta_{2} C_{12'}+p_{2}\eta_{1}\eta_{2}C_{22'})}=\frac{C_{12'}}{p_1C_{12'}+p_2C_{22'}}+1.
\end{equation}
\begin{equation}
g_{22'}=\frac{p_{2}\eta_{i}\eta_{1}\eta_{2} C_{22'}+p_{2}\eta_{i}(p_1\eta_{1}\eta_{2} C_{12'}+p_{2}\eta_{1}\eta_{2}C_{22'})}{p_{2}\eta_{i}(p_1\eta_{1}\eta_{2} C_{12'}+p_{2}\eta_{1}\eta_{2}C_{22'})}=\frac{C_{22'}}{p_1C_{12'}+p_2C_{22'}}+1.
\end{equation}

It can be clearly seen that the cross-correlation function between uncorrelated channels (1 and 2$'$, 2 and 1$'$) will be 1 - caused by accidental coincidences of uncorrelated background - only if no crosstalk exists ($C_{12'}$=0, $C_{21'}$=0), otherwise it will be larger than 1. Here, according to the experimental parameters, we assume $g_{12'}$=1 (i.e., $C_{12'}$=0) and $p_1$=$p_2$=$p$=0.0354, then calculate the crosstalk $C_{21'}$ between different channels with different $g_{21'}$. The calculated result is shown in Fig.S. \ref{fig:5-2}(b). It indicates that the crosstalk between different channels increases with the increasing of cross-correlation function, thus cross-correlation function is a good indicator to characterize the crosstalk.

\subsection*{Note 6: The crosstalk of the source and detection system without quantum memory. }
\begin{figure}[ht]
\centering
\includegraphics[width=7cm]{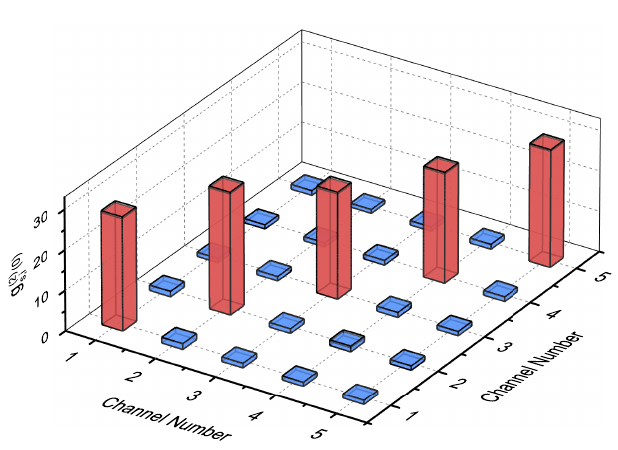}
\caption{5 $\times$ 5 array of $g_{s,i}^{(2)}(0)$ between different frequency channels of the photon-pairs source. All error bars are calculated by standard deviations of counts which obey Poisson distribution.}
\label{fig:6-1}
\end{figure}

\noindent \\The 5 $\times$ 5 array of $g_{s,i}^{(2)}(0)$ between different spectral channels of the photon-pairs source is shown in Fig.S. \ref{fig:6-1}. The results indicate that crosstalk of different spectral channels of the source and detection system is negligible.

\subsection*{Note 7: 1650$\times$1650 array of $g_{s,i}^{(2)}(0)$ between 1650 modes}
\begin{figure}[ht]
\centering
\includegraphics[]{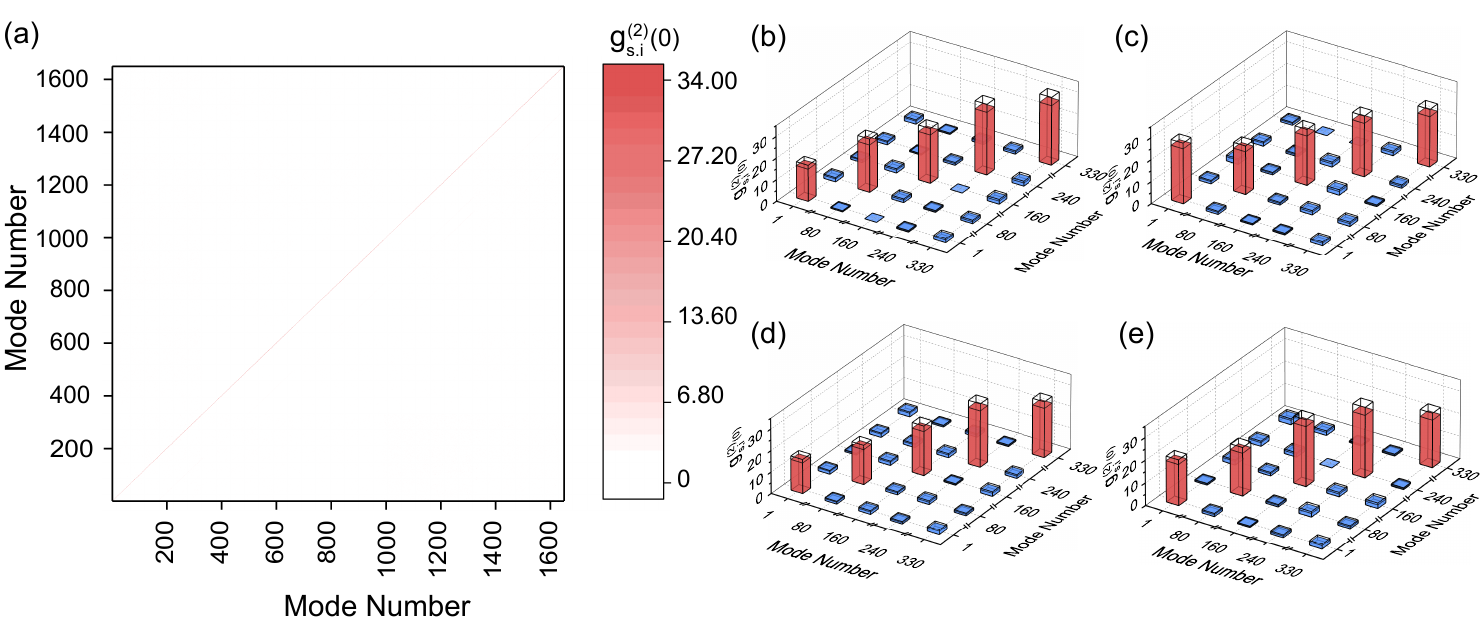}
\caption{(a) 1650$\times$1650 array of $g_{s,i}^{(2)}(0)$ between 1650 modes. (b), (c), (d), (e) Parts of results of crosstalk measurement among different temporal modes retrieved from channel 1, 3, 4, 5. All error bars are calculated by standard deviations of counts which obey Poisson distribution.}
\label{fig:7-1}
\end{figure}

\noindent \\The total number of spectro-temporal modes is 1650, and the cross-correlation function between 1650 single photon modes is shown in Fig.S. \ref{fig:7-1}(a). (b), (c), (d), and (e) in Fig.S. \ref{fig:7-1} present parts of $g_{s,i}^{(2)}(0)$ corresponding to different temporal modes retrieved from channel 1, 3, 4, and 5. (modes numbers: 1, 80, 160, 240, 330).

\clearpage
\subsection*{Note 8: Impedance-matched in-fibre cavity quantum storage}
\begin{figure}[ht]
\centering
\includegraphics[width=10cm]{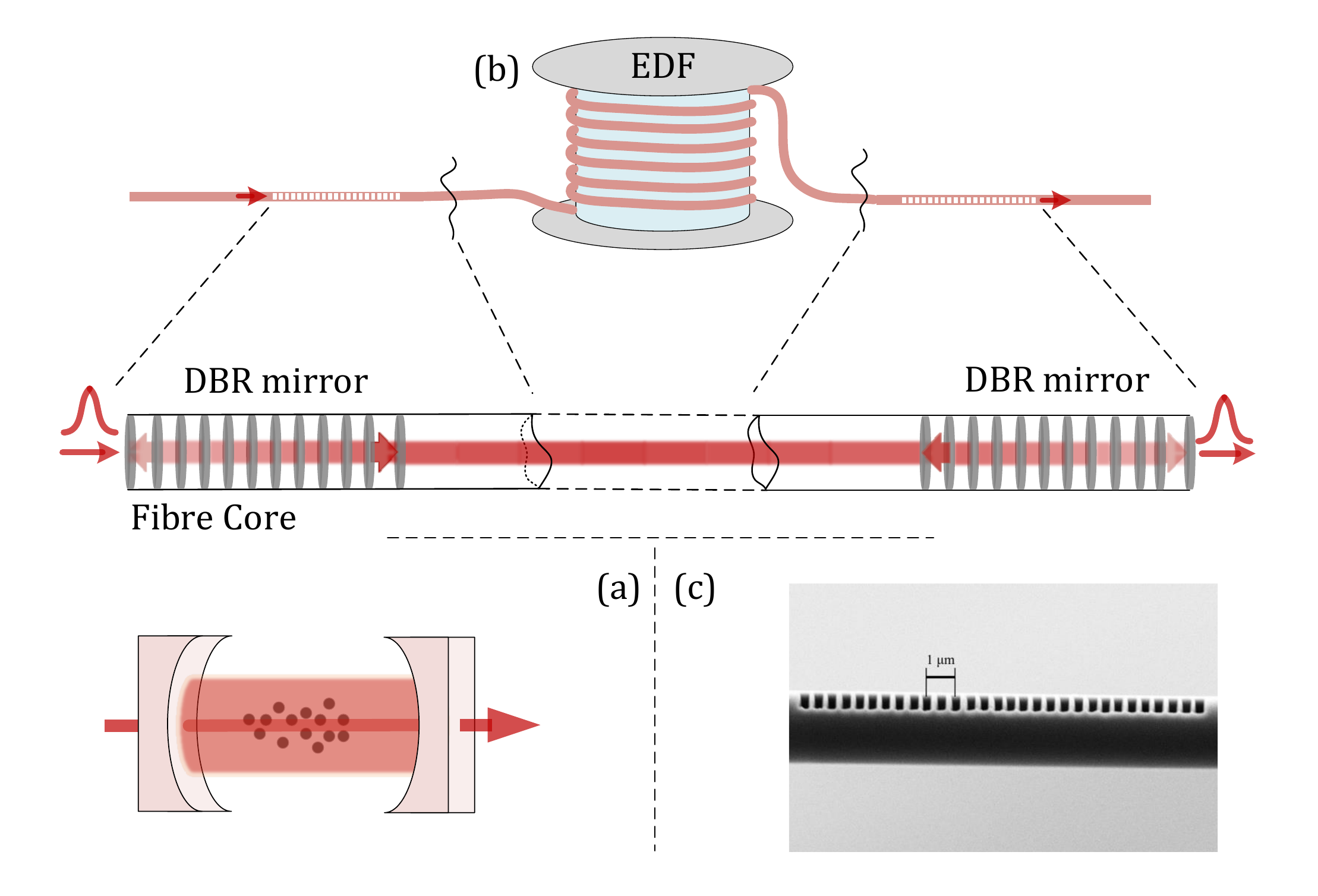}
\caption{Impedance-matched in-fibre cavity quantum storage. (a) Conception of impedance-matched cavity quantum storage. (b) Scheme of EDF based impedance-matched quantum memory. Two DBR mirrors are fabricated at both ends of the EDF. (c) Our recent progress on the fabrication of EDB mirror in fibre.}
\label{fig:8-1}
\end{figure}

\noindent \\We propose a scheme based on a piece of EDF with low doping concentration and an impedance-matched in-fibre cavity to realize a quantum storage system. The Fig.S. \ref{fig:8-1}(a) shows the conception of impedance-matched cavity storage. As shown in Fig.S. \ref{fig:8-1}(b), in this scheme two distributed Bragg reflection (DBR) mirrors are fabricated along both ends of the EDF forming an in-fibre cavity. The photons sent into the in-fibre cavity and subsequently are reflected back and forth through the DBR mirror at both ends, increasing the light-matter interaction, thus achieving high-efficient quantum storage. In addition, the storage time is improved owing to the low doping concentration of erbium ions. Currently, towards this scheme we have fabricated a DBR mirror on fibre with micro/nano fabrication (see Fig.S. \ref{fig:8-1}(c)).

\subsection*{Note 9: Measurement of lifetime of Zeeman sublevels of ground state}
\begin{figure}[ht]
\centering
\includegraphics[width=12cm]{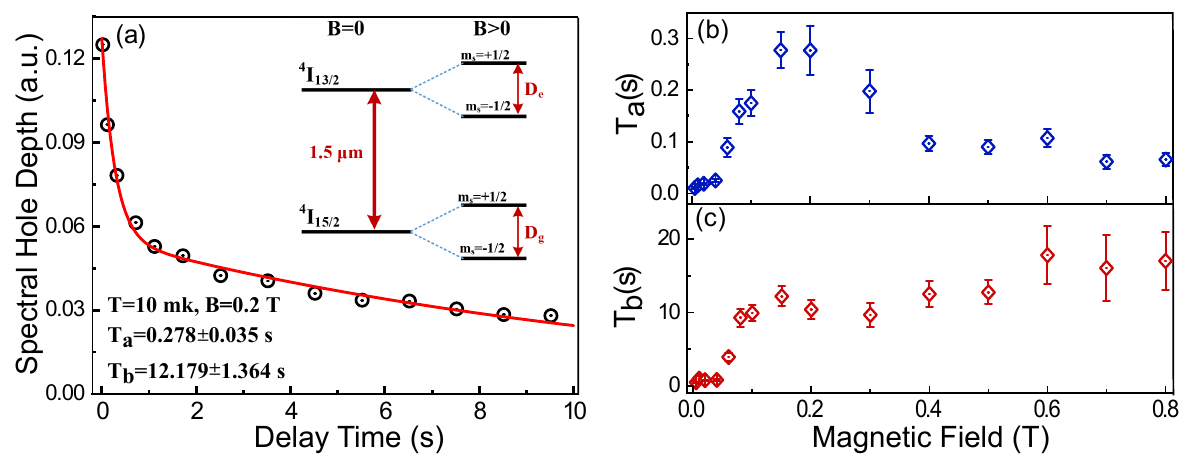}
\caption{Measurement of lifetime of Zeeman sublevels through spectral holes burning. (a) A typical result of the hole decay curve at a temperature of 10 mK and a magnetic field of 0.2 T. Inset: the simple energy levels of the EDF before/after applying the magnetic field. (b) and (c) Lifetime of Zeeman sublevels as a function of applied magnetic field strength (slow and fast decay, respectively).}
\label{fig:9-1}
\end{figure}

\noindent \\The preparation of AFC requires frequency-selective population transfer from the electronic ground level through the excited level into an auxiliary level, i.e., long-lived Zeeman sublevels. The lifetime of the Zeeman sublevels determines the efficiency of optical pumping. Moreover, long-lived Zeeman sublevels also determine the persistent time of AFC. To acquire the lifetime of Zeeman sublevels of our EDF, we observe the decay of persistent holes with the delay time between burning and probing pulses under magnetic fields from 0 T to 0.8 T. The Fig.S. \ref{fig:9-1}(a) shows a typical result of the hole decay curve. All decay curves can be fitted by a double exponential function, $I(t)=I_ae^{-t/T_a}+I_be^{-t/T_b}$, where $t$ is the time delay between burning and probing pulses, $T_a$ and $T_b$ are the lifetimes of two different holes, and $I_a$ and $I_b$  are two initial amplitudes at $t=0$. From the fitting, it is likely that two different classes of erbium ions could be observed as depicted in previous work14. The fast decay time ($T_a$) of $\sim$ 100 ms and the slow decay time ($T_b$) of $\sim$ 10 s are measured with different values of magnetic fields (as shown in Fig.S \ref{fig:9-1}(b) and (c)). With a magnetic field of 0.2 T, $T_ a$ reaches 0.278 $\pm$ 0.035 s, therefore, in our experiment we choose the magnetic field of 0.2 T.

\subsection*{Note 10: Selection of the waiting time before storing signal photons}
\begin{figure}[ht]
\centering
\includegraphics[width=6.5 cm]{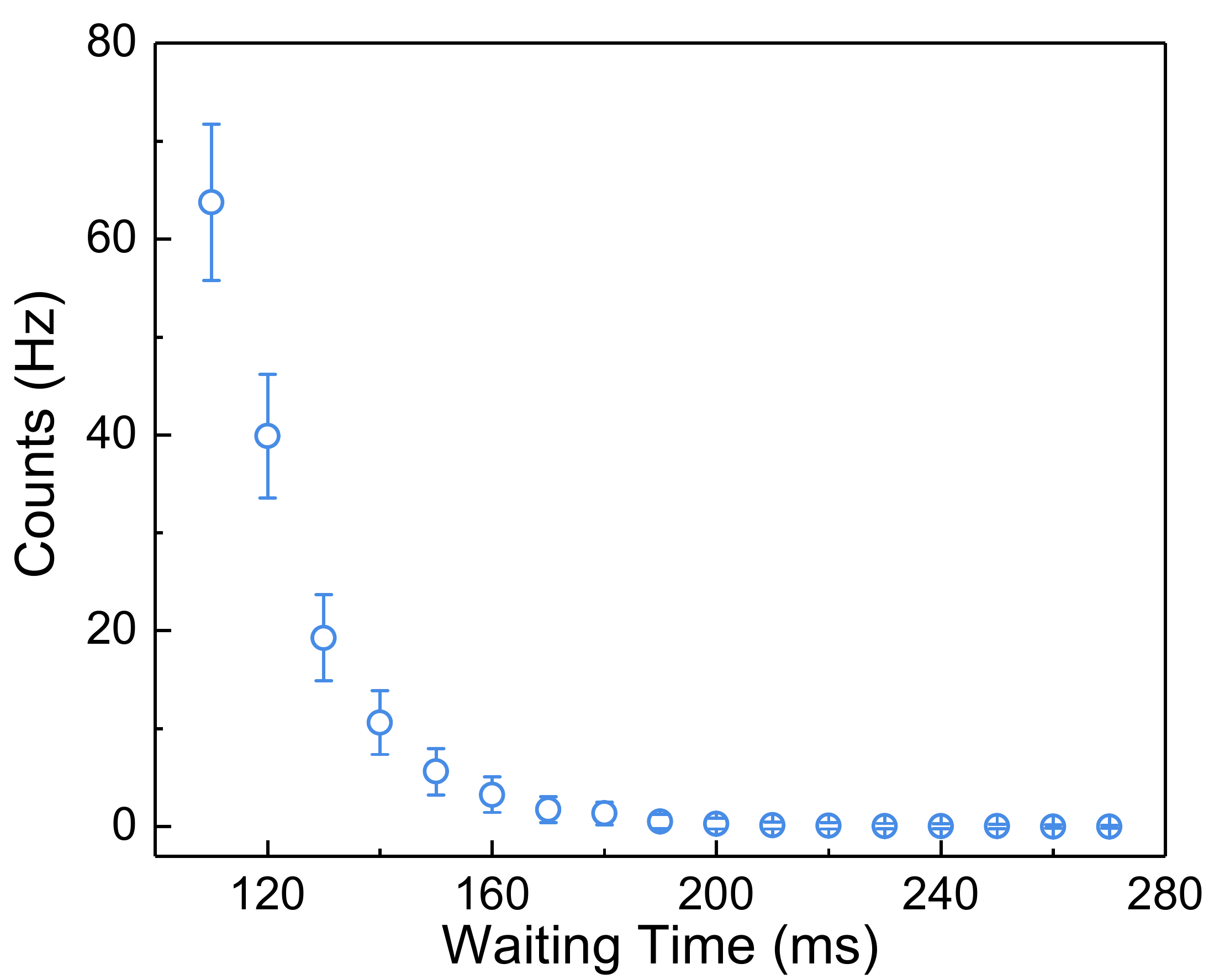}
\caption{Counts of spontaneously emitted photons in 1-ns-wide time window versus waiting time. All error bars are calculated by standard deviations of counts which obey Poisson distribution.}
\label{fig:10-1}
\end{figure}

\noindent \\In order to ensure that recalled photons are not infected by spontaneously emitted photons after spectral tailoring, we measure the counts of spontaneously emitted photons in 1-ns-wide time window versus waiting time. The measuring results are shown in Fig.S. \ref{fig:10-1}. According to the measured results, the waiting time is set to 200 ms.

\clearpage

\end{document}